\documentclass[10pt,aps,prl,twocolumn,superscriptaddress]{revtex4-1}
\pdfoutput=1 
\usepackage{microtype}
\usepackage{times}
\usepackage{amsfonts}
\usepackage{mathrsfs}
\usepackage{amsmath}
\usepackage{amssymb}
\usepackage{fancyhdr}
\usepackage{graphicx}
\usepackage{xspace}
\usepackage{rotating}
\usepackage[normalem]{ulem}
\usepackage{braket}
\usepackage{verbatim}
\usepackage{xcolor}
\usepackage[utf8]{inputenc}
\usepackage[medium]{titlesec}
\usepackage{bm}
\usepackage[normalem]{ulem}
\usepackage{extarrows}
\usepackage{slashed}
\usepackage{isodateo}
\usepackage{graphicx}
\usepackage{xcolor}
\usepackage[bookmarksnumbered=true,bookmarksopen=true]{hyperref}
\usepackage[hmargin=.7in,vmargin=1.1in]{geometry}
\usepackage{indentfirst}
\usepackage{cancel}

\graphicspath{{./}{./Figs/}{./figs/}{../Figs/}{./fig/}}

\usepackage{amsfonts}
\usepackage{mathtools}
\usepackage{pifont}
\usepackage{mathrsfs}
\usepackage{amsmath}
\usepackage{amssymb}
\usepackage{framed}

\newcommand{\nc}{\newcommand}

\nc{\beqa}{\begin{eqnarray}}  
\nc{\eeqa}{\end{eqnarray}}  
\nc{\bit}{\begin{itemize}}  
\nc{\eit}{\end{itemize}}

\newcommand{\benum}{\begin{enumerate}}
\newcommand{\eenum}{\end{enumerate}}
\newcommand{\bi}{\begin{itemize}}
\newcommand{\ei}{\end{itemize}}
\newcommand{\Mp}{M_{\rm pl}}

\newcommand{\beq}{\begin{equation}}
\newcommand{\eeq}{\end{equation}}

\newcommand{\bea}{\begin{eqnarray}}
\newcommand{\eea}{\end{eqnarray}}

\newcommand{\Rmnum}[1]{\expandafter\@slowromancap\romannumeral #1@}

\def\dx{\mathrm{d}}

\def\bga{\begin{aligned}}
\def\eda{\end{aligned}}
\def\bgp{\begin{pmatrix}}
\def\edp{\end{pmatrix}}
\def\bgs{\begin{subequations}}
\def\eds{\end{subequations}}

\def\to{\rightarrow}

\def\Mp{M_{\text{Pl}}}

\renewcommand{\rm}{\mathrm}

\definecolor{cardinal}{rgb}{0.77, 0.12, 0.23}

\usepackage{eso-pic}
\newcommand{\reportnum}[2]{
  \AddToShipoutPictureBG*{%
    \AtPageUpperLeft{%
      \hspace{0.75\paperwidth}%
      \raisebox{#1\baselineskip}{%
        \makebox[0pt][l]{\textnormal{#2}}
  }}}%
}

\begin{document} 

\reportnum{-3}{KEK-TH-2634}
\reportnum{-4}{KEK-Cosmo-0349}

\title{Gravitational Wave Symphony from Oscillating Spectator Scalar Fields}

\author{Yanou Cui}
\email{yanou.cui@ucr.edu}
\affiliation{Department of Physics and Astronomy, University of California, Riverside, California 92521, USA}

\author{Pankaj Saha}
\email{pankaj@post.kek.jp}
\affiliation{Institute of Particle and Nuclear Studies~(IPNS), High Energy Accelerator
Research Organization (KEK), Oho 1-1, Tsukuba 305-0801, Japan}
\affiliation{School of Natural Sciences, 
Seoul National University of Science and Technology,\\ Seoul 01811, Republic of Korea}

\author{Evangelos I. Sfakianakis}
\email{evangelos.sfakianakis@case.edu}
\affiliation{Institut de F\'isica d'Altes Energies (IFAE), The Barcelona Institute of Science and Technology (BIST), Campus UAB, 08193 Bellaterra, Barcelona, Spain}
\affiliation{Department of Physics, Case Western Reserve University, Cleveland, Ohio 44106, USA}

\date{\today}

\begin{abstract}
We investigate a generic source of stochastic gravitational wave background due to the parametric resonance of oscillating scalar fields in the early Universe. By systematically analyzing benchmark models through lattice simulations and considering a wide range of parameters, we demonstrate that such a scenario can lead to detectable signals in gravitational wave detectors over a broad frequency range and potentially address the recent findings by pulsar timing array experiments. Furthermore, these models naturally yield ultralight dark matter candidates or dark radiation detectable by cosmic microwave background observatories.


\end{abstract}
\maketitle

\noindent \textit{Introduction.} 
The discovery of gravitational waves (GWs) by the LIGO-Virgo collaboration in 2015 \cite{LIGOScientific:2016aoc} heralded the beginning of a new era of observational astronomy, as well as unprecedented opportunities for new discovery in particle physics and cosmology.
The very recent compelling evidence for a stochastic GW background (SGWB) reported by the pulsar timing array (PTA) experiments such as NANOGrav~\cite{NANOGrav:2023gor}, EPTA+InPTA~\cite{EPTA:2023fyk, EPTA:2023xxk}, PPTA\cite{Reardon:2023gzh}, and CPTA\cite{Xu:2023wog} with plausible cosmological origins \cite{NANOGrav:2023hvm}, further propels the emerging research field of using SGWB to probe the pre-Big Bang Nucleosynthesis \textit{``primordial dark age''} \cite{Boyle:2005se,Boyle:2007zx,Cui:2017ufi,Cui:2018rwi,Caldwell:2018giq} and the associated fundamental physics. Among the well-motivated sources of cosmogenic SGWBs, a few classes have been widely and systematically studied, such as those originating from inflation, cosmic strings, and phase transitions \cite{Caprini:2018mtu,Caldwell:2022qsj}. Another generic source for SGWBs is parametric resonance excited during the oscillation of a scalar field, leading to rapid, nonperturbative particle production and thus large time-dependent field inhomogeneities \cite{Kitajima:2018zco,Lozanov:2019ylm,Hiramatsu:2020obh, Zhou:2020kkf}. Existing studies in this direction mostly focus on specific scenarios, such as those related to inflationary preheating or Affleck-Dine baryogenesis \cite{Easther:2006vd,Easther:2006gt,Garcia-Bellido:2007nns,Garcia-Bellido:2007fiu,Dufaux:2007pt,Easther:2007vj,Ghoshal:2022jdt,White:2021hwi}. Despite being highly motivated, the additional constraints associated with these models (e.g.~on inflation or low-energy supersymmetry breaking scale) limit the detectability of their GW signals: either the frequency is too high, or the signal strength is too small, for observations in foreseeable future (for exceptions, e.g., by utilizing axion-gauge or axion-gravity coupling, see~\cite{Cui:2021are, Machado:2018nqk, Madge:2021abk, Li:2023vuu}).

In this Letter, we systematically expand the studies on GWs sourced through parametric resonance by considering a generic spectator scalar field displaced from its true vacuum during inflation. 
After inflation, it rolls toward its potential minimum and oscillates coherently until (some of) its energy dissipates into particle excitations and GWs. We investigate simple, minimal benchmark models with renormalizable scalar field potentials, allowing model parameters (masses and couplings) to vary over a wide range. %
We found 
detectable SGWB signals with  peak frequency lying in a wide range from ${\cal O}(0.01)$ nHz to GHz,
and  peak signal strength up to $ \Omega_{\rm {GW}}\sim 10^{-9}$.
Notably, when the (effective) mass of the scalar field is ${\cal O}(10^{-13}) $ eV, this class of models may provide another plausible cosmological explanation for the recent observation by pulsar timing arrays, complementing the existing literature on this timely topic. 
Interestingly, in such PTA-relevant models, the massive scalar field also serves as an ultralight dark matter (DM) candidate. 
Furthermore, models involving massless scalar field(s), a form of dark radiation, potentially lead to $\Delta N_{\rm{eff}}$ observable with upcoming cosmic microwave background (CMB) and large-scale structure experiments. Therefore, our study 
demonstrates an intriguing complementarity between GW signal and dark matter or CMB physics that generically emerges from these types of models. 

The rest of the \textit{Letter} is organized as follows. In the next section, we present benchmark models to be examined, along with analytic expectations for GW production. 
Following that, we show the numerical results for GW signals and the complementary observables (DM and/or $N_{\rm{eff}}$).
We conclude with an outlook.

\medskip

\noindent \textit{Benchmark Models and Estimates for GW Signals.}
The equation of motion for a scalar field oscillating in an Friedman-Robertson-Walker background is given by
\beq
\label{eq:phi}
\ddot\phi + 3 H \dot \phi-\frac{1}{a^2}\nabla^2 \phi +\frac{\partial V}{\partial \phi }=0\, .
\eeq
For a massive scalar field, the last term is $\partial V/\partial \phi \supset m_\phi^2 \phi$. 
For the remainder of this Letter, we assume that the Universe is radiation dominated throughout the time range of interest, i.e. $\rho_\phi$ is a subleading component, $\rho_\phi < 3 H^2M_{\rm {Pl}}^2$, and $H=1/2t$ ($M_{\rm {Pl}}$ is the Plank scale). 
When $H\gg m_\phi$, the field is frozen and only starts rolling or oscillating significantly after $H_{\rm osc}\sim m_\phi$. For a massless scalar field with $V\supset \lambda \phi^4$, the field is again frozen at early times, when $H$ dominates over $V_{,\phi\phi}$, and starts rolling appreciably when $H_{\rm osc}\simeq \sqrt{\left | V_{,\phi}/\phi \right |} = \sqrt{\lambda_\phi} \phi_{\rm {in}}$, where $\phi_{\rm {in}}$ is the initial field amplitude (see, e.g.,~\cite{Gasparotto:2023psh}).
We take the initial field displacement, $\phi_{\rm {in}}$, as set during inflation. It typically needs to be up to $\sim M_{\rm {Pl}}$ for sufficient GW signal, as we found, which can be realized in various ways. First, it could be acquired due to quantum fluctuations during inflation, where $\phi_{\rm {in}}\sim M_{\rm {Pl}}$ requires many e-folds of inflation ($N$)  and a small quartic coupling $\lambda$ in the case of a self-interacting $\phi$ (see Supplemental Material for a detailed derivation\cite{cite:sup}). 
Second, such a large displacement can be realized by introducing nonrenormalizable terms in the potential, which shift the vacuum expectation value of $\phi$ during inflation \cite{Dine:1995kz}. Finally, a large $\phi_{\rm {in}}$ could be simply due to an \textit{ad hoc} initial condition, similar to the case of the inflaton. Our result is independent of the specifics of how $\phi_{\rm {in}}$ is realized.

We consider the possibilities of both a massive and massless scalar field $\phi$ and choose two different models in each case, including only renormalizable terms (up to quartic). These can be summarized as follows 
\begin{align}
\text{Model A:}\quad~V &=  {m^2_\phi \over 2}\phi^2 + {g\over 2} \phi^2\chi^2,\\
\text{Model B:}\quad~V &=  {m^2_\phi \over 2}\phi^2 + {\lambda_\chi \over 4}\chi^4 + {\sigma\over 2} \phi\chi^2,\\
\text{Model C:}\quad~V &=  {\lambda \over 4}\phi^4,\\
\text{Model D:}\quad~V &=  {\lambda \over 4}\phi^4 + {g\over 2} \phi^2\chi^2,
\end{align}
In models A and B, a free massive scalar $\phi$ couples to a massless scalar $\chi$ through a quartic (A) or trilinear (B) term. The second scalar field $\chi$ is necessary in this case, as a free massive scalar does not exhibit parametric self-resonance. In contrast, for a self-interacting scalar, such a coupling is not necessary, and the quartic scalar potential is sufficient to generate parametric resonance and GWs (C). Finally 
we consider a massless quartic field $\phi$ coupled to a massless field $\chi$ through a quartic interaction (D).
In light of the constraints related to BBN, we require that $H_{\rm {osc}}>H_{\mathrm{BBN}}$, leading to $m_\phi \gtrsim 10^{-16}\,{\rm {eV}} $ and $\sqrt{\lambda_\phi}\phi_{\rm {in}}\gtrsim10^{-16}\,{\rm {eV}} $ for the massive and massless cases, respectively. The lower bound of $m_\phi\sim10^{-16}$ eV corresponds to a GW frequency of ${\cal O}(0.01)$ nHz.

In order to maximize the efficiency of parametric resonance, we keep the $\chi$ field VEV at zero. During inflation, the quantum fluctuations of $\chi$ can be suppressed by the coupling to the inflaton, stabilizing it at the origin with a Hubble-scale mass~\cite{Dine:1995kz}. Even if $\chi$ acquires a VEV during inflation, a coupling to the thermal plasma after reheating can lead to a large thermal mass that dynamically drives the VEV to zero. This provides interesting avenues for complementarity studies by considering various couplings of the $\chi$ to the standard model (SM), which we leave for future work. 

For a simpler classification of the models into massive and massless and to minimize the number of model parameters, we did not include combinations such as $\lambda_\phi \phi^4/4 + m_\phi^2\phi^2/2$. This does not reduce the general applicability of our models for the following reasons. If, during the rolling of the $\phi$ field, the motion is dominated by the quadratic term,  the system is well-described by models A and B. If the two terms are comparable at first, the first few oscillations will be anharmonic, but the quartic term would quickly become subdominant to the quadratic. If, alternatively, the quadratic term is negligible relative to the quartic, we can follow the evolution of models C and D until the mass becomes important and the $\phi$ particles become nonrelativistic. It is straightforward to adapt our analysis to include these variations.

We separate the field $\phi$ as $\phi(x,t) = \phi_{\rm {bg}}(t) 
 + \delta\phi(x,t)$, where $ \phi_{\rm {bg}}(t) $ is the background value, which follows Eq.~\eqref{eq:phi} with $\nabla^2\phi_{\rm {bg}}=0$. The fluctuation component can be expanded in Fourier modes as
 \beq
\delta\phi(x,t) = \int {d^3\over (2\pi)^3} \left[
a_k e^{i\vec k \cdot \vec x} \delta\phi_k(t)+a_k^\dagger e^{-i\vec k \cdot \vec x} \delta\phi^*_k(t)
\right ]
 \eeq
 The $\chi$ field background value is assumed to be 0, thus $\chi(x,t) = \delta\chi(x,t)$, and can be similarly decomposed into modes $\delta \chi_k(t)$.

As $\phi_{\rm {bg}}$ oscillates around the minimum of its potential, the fluctuations  $\delta\phi$ and $\delta\chi$ exhibit parametric resonance~\cite{Figueroa:2016wxr}, similar to preheating following inflation~\cite{Kofman:1997yn, Greene:1997fu}. At the linear level, one needs to solve Eq.~\eqref{eq:phi} for $\phi_{\rm {bg}}$ (with $\nabla^2\phi_{\rm {bg}}=0$), and the equations of motion of the fluctuation are
\begin{eqnarray}
\label{eq:phifluct}
\ddot\delta\phi_k + 3H\dot{\delta\phi_k} + {k^2\over a^2}\delta\phi_k + \left .{\partial^2 V\over \partial\phi^2}\right|_{\phi_{\rm {bg}}}\delta\phi_k=0 \\
\label{eq:chifluct}
\ddot\delta\chi_k + 3H\dot{\delta\chi_k} + {k^2\over a^2}\delta\chi_k + \left . {\partial^2 V\over \partial\chi^2}\right|_{\phi_{\rm {bg}}}\delta\chi_k=0 
\end{eqnarray}
Parametric resonance amplifies a certain range of wave numbers, typically up to ${\cal O}(m_\phi)$.
We can thus approximate the spectrum of scalar fluctuations as peaking
around a characteristic scale $k_c$, which depends on model parameters and is proportional to $H_{\rm {osc}}$.

GW production from peaked sources can be estimated by the rule of thumb as given in~\cite{Giblin:2014gra}
\begin{eqnarray}\label{eq:freq}
\nu_{\rm {GW}}^{\rm {peak}} &=& 2.7\times 10^{10} \sqrt{\frac{H_c}{\Mp}}\frac{k_c}{H_c}\, {\rm {Hz}} \, ,
\\
\Omega_{\rm {GW}}^{\rm {peak}} &=& 2.3\times 10^{-4} \, \alpha^2\, \beta\, w 
\left ( {k_c\over \sigma} \right ) 
\left ({H_c \over k_c}\right )^2.
\label{eq:Omega}
\end{eqnarray}
In these formulas, $\alpha$ is the fraction of the energy in the GW source (the amplified scalar fluctuations in our case) relative to the total energy density of the Universe, and $\beta$ encodes the anisotropy of the source, which can be extracted from simulations, with $\beta={\cal O}(0.01-0.1)$ as a reasonable range \cite{Giblin:2010sp}.
The equation of state of the Universe at the time of GW emission is parametrized by $w=1/3$ for production during the radiation-dominated era.
$\sigma$ is the width of the source in momentum space, which is estimated to be $\sigma\sim k_c$ for peaked sources. 
For a massive field, we typically get $k_c\sim m_{\phi} \simeq H_{\rm {osc}} $. However, the Hubble scale during GW emission $H_{c}$ (close to the end of parametric resonance) is smaller than $H_{\rm {osc}}$ due to the expansion of the Universe during the span of the parametric resonance era. The exact ratio $H_{\rm {osc}}/ H_{c}$ is model-dependent but can be as large as $10^4$ as per our numerical finding. Based on our simulations, we found $\nu_{\rm{GW}}^{\rm{peak}} \sim {\cal O}(10^{12}) \sqrt{H_{\rm {osc}}/M_{\rm {Pl}}} $ Hz to be an accurate estimate. For a massless field, our simulation results agree with Eq.~\eqref{eq:freq} taking $H_c = H_{\rm {osc}}$. 
Note that $\nu_{\rm {GW}}^{\rm {peak}}$ includes the redshift after production, which depends on thermal history. We assume the standard radiation-dominated for concreteness, while a prolonged matter-dominated phase would modify our results. 

The transverse-traceless tensor perturbations $h_{ij}$ follows the linearized equation
\begin{align}
    \ddot{h}_{ij} + 3H\dot{h}_{ij} - \frac{\nabla^2h_{ij}}{a^2} = \frac{2}{\Mp^2 a^2}\Pi_{ij}^{TT},
    \label{eq:eqhij}
\end{align}
where $\Pi_{ij}^{TT}$ represents the transverse-traceless part of the effective anisotropic stress tensor $\Pi_{ij}= \partial_{i}\phi\partial_{j}\phi + \partial_{i}\chi\partial_{j}\chi$. 
The quantity of interest is the energy density power spectrum of the GWs normalized by the critical energy density of the Universe today, $\rho_c$. 
\beq
\Omega_{\mathrm{GW}}(k,t) = \frac{1}{\rho_c}\frac{d\rho_{\mathrm{GW}}}{d\log k}=\frac{k^3}{(4\pi)^3GV}\int\frac{d\Omega_k}{4\pi}\dot{h}_{ij}\dot{h}^{\ast}_{ij}
\eeq
where the differential $d\Omega_k$ is the solid angle element in Fourier space and the integral is over the Fourier modes $h_{ij}(\mathbf{k},t)$. The simulations were performed using
 the publicly available code $\bm{\mathcal{C}\texttt{osmo}\mathcal{L}\texttt{attice}}$~\cite{Figueroa:2020rrl,Figueroa:2021yhd} (the numerical setup and convergence tests are presented in Sec. B.1, B.4 of the Supplemental Material~\cite{cite:sup})

\medskip
\par
\begin{figure}
\includegraphics[scale=0.4]{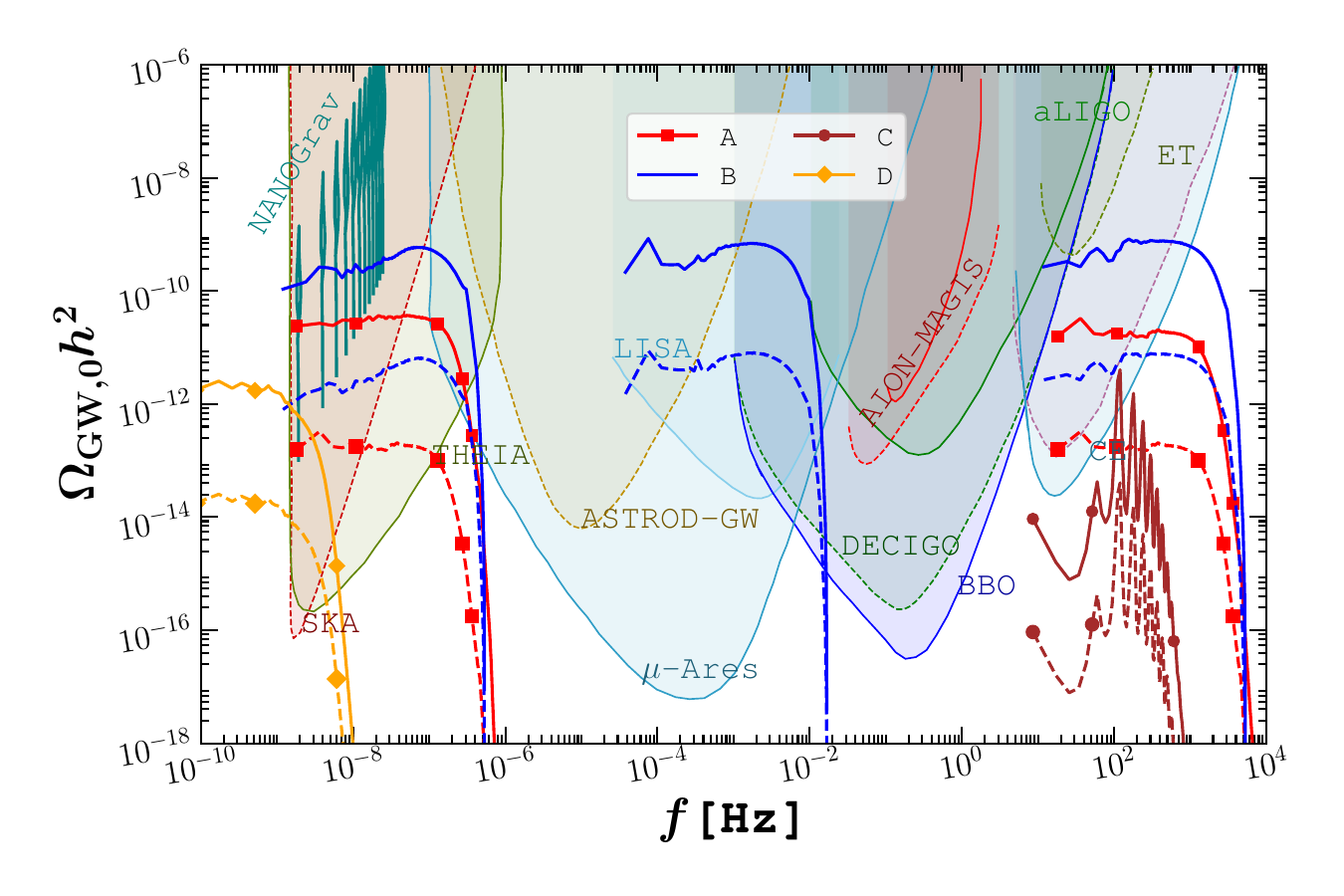}
\caption{Example results for Models A, B, C, and D (red, blue, brown, and orange), corresponding to the parameter choices given in Table~\ref{tab1}. Solid and dashed lines indicate the results when the initial energy in the scalar is assumed to be 10\% and 1\% of the total energy density of the Universe, respectively. Note that model C has multipeaked oscillatory features. The constraints and sensitivities of various GW experiments are also shown.} 
\label{fig:omega_freq}
\end{figure}

\noindent
\textit{Numerical results of the GW signals and complementary phenomenology: Dark matter and dark radiation. } In Fig.~\ref{fig:omega_freq} and Table~\ref{tab1}, we illustrate the numerical results from our simulation for a set of benchmark models, which demonstrate detectable GW signals over a wide frequency range. In the following, we elaborate on the relevant dynamics 
to provide physics insights for interpreting these results.

For the massive models A and B,
~$\phi_{\rm {bg}}$ starts rolling and oscillating when $H_{\rm {osc}}\simeq m_\phi$, and the ratio of the energy density in the $\phi$ field over the total energy density [roughly equals the parameter $\alpha$ in Eq.~\eqref{eq:Omega}] is $\rho_\phi / \rho_{\rm {tot}} = (m_\phi^2 \phi_{\rm {in}}^2 /2)/(3H_*^2M_{\rm {Pl}}^2)\simeq \phi_{\rm {in}}^2/(6M_{\rm {Pl}}^2) $, leading to, e.g.,~$\phi_{\rm {in}}/M_{\rm {Pl}} = 0.77,0.25$ for $10$\%, $1$\% ratio, respectively.
\\
\indent In order to get sufficient GW signals, the transfer of a significant fraction of $\rho_\phi$ into field fluctuations is required. The modes $\delta\chi_k$ undergo parametric resonance since the interaction with the inflaton or thermal bath that stabilized $\chi$ at zero becomes inactive at the late times when $\phi$ oscillation starts.
Because of rescattering effects, $\delta\phi_k$ modes will also be generated; however, they can be at most comparable to $\delta\chi_k$ in power. Thus, we can safely neglect the $\delta\phi_k$ modes in our analytic estimates (but included in the numerical simulation).
By measuring $t, k$ in units of ${m_\phi}$, the last term of Eq.~\eqref{eq:chifluct} becomes $q_{A,B} f(t) \delta\chi_k$, where 
$q_A\equiv g\phi_{\rm {in}}^2/m_\phi^2$ and $q_B = \sigma \phi_{\rm {in}}/m_\phi^2$ are dimensionless quantities controlling the strength of parametric resonance in models A and B, respectively.
$f(t)$ is a decaying periodic function, bounded by $\pm 1$ and encodes the oscillating background field $\phi_{\rm {bg}}$.
\\
\indent The dimensionless form of Eq.~\eqref{eq:chifluct} indicates that during parametric resonance the amplification of $\delta\chi_k$ is independent of $m_\phi$ and evolves as $\delta\chi_k\propto e^{\mu_k t}$, where $\mu_k$ is called the Floquet exponent.
After a time $t$, the ratio of the energy density of $\delta \chi$ modes to the $\phi$ energy density scales as $\rho_{\delta\chi}/ \rho_\phi\sim H^4e^{2\mu t} / H^2M_{\rm {Pl}}^2 \sim (m_\phi e^{\mu t}/M_{\rm {Pl}})^2 $, since we take $\rho_\phi$ to be a fraction of the total energy of the Universe and $H\sim m_\phi$ during the (early stages of) parametric resonance. We thus see that the amplification required to achieve a similar energy transfer efficiency must be larger for smaller values of $m_\phi$. By measuring both $\mu$ and time in units of $m_\phi$, 
 we deduce that for smaller values of $m_\phi$, parametric resonance needs to be active for more oscillation periods of the background field $\phi_{\rm {bg}}$, which agrees with 
 our numerical simulations. Consequently, the endvalue of $\rho_{\delta\chi}/\rho_{\rm {tot}}$ is largely independent of $m_\phi$; thus we expect $\Omega_{\rm {GW}}$ also to be $m_\phi$-independent while $\nu_{\rm {GW}}$ can vary without affecting $\Omega_{\rm {GW}}$~\cite{Note1}.
 In reality, there is a mild dependence on $m_\phi$, due to the different redshift behaviors of matter and radiation during the prolonged parametric resonance, which manifests itself as a small suppression of $\Omega_{\rm {GW}}$ for nHz frequencies 
(see Sec. B.2, B.3 of the Supplemental Material for more details~\cite{cite:sup}).

Models C and D exhibit similar dynamics as models A and B. The $\phi$ field starts rolling or oscillating at $H_{\rm {osc}} \simeq \sqrt{\lambda_\phi}\phi_{\rm {in}}$, whereas the initial field amplitude is given by the energy ratio $\rho_\phi / \rho_{\rm {tot}} = (\lambda_\phi \phi^4_{\rm {in}}/4) / (3M_{\rm {Pl}}^2 H_{\rm {osc}}^2) \simeq \phi_{\rm {in}}^2/(12 M_{\rm {Pl}}^2)$. The initial field amplitude is $\phi_{\rm {in}}/M_{\rm {Pl}} = 1.17, 0.35$ for $10\%$, $1\%$ energy ratio. 
In model C, parametric resonance amplifies  $\delta\phi$ fluctuations without the need for a second field $\chi$. 
The resulting GW spectrum exhibits a multipeak structure, which is absent in all other models and provides a distinct feature. The origin of the multipeak structure in model C can be best understood by comparing it to the similar model D. We see that these features are accompanied by an overall reduction in the peak GW amplitude. Self-resonance in model C is weaker than the parametric resonance of $\chi$ in model D (and similarly in models A and B). A weak resonance amplifies specific waven umbers instead of a broad range of modes, leading to a multipeak structure, which would be washed out for more efficient parametric resonance. A similar effect appears in oscillon preheating, where strong resonance washes out multi-peak features in the GW signal \cite{Hiramatsu:2020obh}.

We find that for all the models we consider, in order to produce sufficient GW signals for detection, the initial displacement $\phi_{\rm {in}}$ is typically driven to be close to $M_{\rm{Pl}}$ (see $\phi_{\rm{in}}$ dependence of $\rho_\phi / \rho_{\rm {tot}}$). This requires a small self-coupling for models $C, D$ in order to yield $\nu_{\rm {GW}}$ in the observable range.
Similarly, for models A and B, the dimensionless couplings between the fields, $g$ or $\sigma/m_\phi$, also must be small.
Since the resonance parameters must be $q_A,q_B \sim{\cal O}(10-10^3)$ for efficient parametric resonance,  $g\sim m_\phi^2/M_{\rm {Pl}}^2\ll 1$ and $\sigma / m_\phi \sim m_\phi / M_{\rm {Pl}}\ll 1$.
These small dimensionless couplings, as preferred, are nevertheless technically natural and thus innocuous theoretically. Furthermore, certain UV completions may provide even more satisfying ways to address the naturalness of such small couplings. For example, the quartic self-coupling of $\phi$ may arise from a loop diagram involving Yukawa coupling of $\phi$ to heavy intermediate sterile fermions. A moderately small Yukawa coupling of values as present in the SM, along with the loop factors and mass suppression from the heavy fermion, can readily yield the very small $\phi$ couplings of our interest. Alternatively, the preferred small quartic couplings may naturally arise from exponential factors (with originally mild hierarchy in parameters) in the framework of a deformed CFT with a dual AdS interpretation, where the scalar is a pseudo-Goldstone boson (dilaton or radion), as suggested in \cite{Agrawal:2016ubh}. Note that by identifying $\phi$ as a Goldstone boson or embedding it in a supersymmetric framework, the concern about radiative stability of $m_\phi$, which is generically present for scalar fields including the SM Higgs, can also be addressed. Nevertheless, the detailed embedding into UV realizations to address naturalness considerations is beyond the scope of this Letter, which focuses on the intriguing phenomenology arising from generic effective field theory models.
\bigskip 

Next, we discuss other phenomenological aspects that naturally complement the GW signals originating from these models. A particularly intriguing possibility arises from models A and B involving a massive scalar field: the scalar field itself can be identified as dark matter. 
Let us consider the state of the scalar field at the time $t_{\rm {end}}$ when parametric resonance production of $\chi$ particles ends, and $t_{\rm {end}}/t_{\rm {osc}} \sim 10^3-10^5$, depending on $m_\phi$.
The energy density of the $\phi$ condensate at $t_{\rm {end}}$ is 
$
\rho_{\phi,{\rm {end}}} = {1\over 2}m_\phi^2\phi_{\rm {end}}^2
$
and afterward redshifts as matter, i.e., $\rho_\phi =\rho_{\phi,{\rm {end}}} (a_{\rm {end}}/a)^3$ ($a$: scale factor). 
Numerically we find e.q.~$\phi_{\rm {end}}\sim{\cal O}(10^{-5})\phi_{\rm {in}} $ for efficient parametric resonance in model A.
By entropy conservation, $g_* a^3T^3$ is  constant ($T$: cosmic temperature). The relic abundance of $\phi$ particles today can then be estimated as
\beq
\Omega_{\phi,0} \equiv {\rho_{\phi,0}\over \rho_{{\rm {tot}},0}}
= {{1\over 2}m_\phi^2 \phi_{\rm {end}}^2 \over 3M_{\rm {Pl}}^2H_0^2} {g_{*,0}\over g_{*,{\rm {end}}}} \left (
T_0 \over T_{\rm {end}}
\right )^3
\, ,
\label{eq:DM}
\eeq
where the subscripts $0$ denote current-day values, we have assumed that the $\phi$ field is stable, in particular, unable to decay to SM particles. This stability assumption is consistent with the models presented and can be reinforced with symmetry protection. 
For model A, $\Omega_{\phi,0}$ monotonically increases with $m_\phi$, and for $m_\phi\gtrsim 10^{-13}$ eV it is larger than the observed $\Omega_{\rm {DM}}\approx 0.2$, causing overclosure problem. Nevertheless, an interesting, viable DM scenario emerges for
ultralight $\phi$ with $m_\phi\sim 10^{-13}$ eV, which also produces a signal in the nHz band in addition to $\Omega_{\phi,0}\sim \Omega_{\rm {DM}}$. 
Meanwhile, the larger $m_\phi$ range for model A, e.g., those relevant for LISA and LIGO ($m_\phi \sim 10^{-2}\, {\rm {eV}}$ for $\nu_{\rm {GW}}\sim 10^{-3}\,{\rm {Hz}}$ and $ m_\phi\sim 10^{6}\, {\rm {eV}}$ for $\nu_{\rm {GW}}\sim 10\,{\rm {Hz}}$), can be viable if $\phi$ is not DM, e.g., by introducing a coupling of $\phi$ to sterile neutrinos as a minimal extension of the model, thus allowing $\phi$ to decay into radiation modes (possibly SM neutrinos at the final stage). 

Unlike model A, model B allows for $\phi$  decay into  pairs of $\chi$ particles with a decay rate $\Gamma_{\phi\to 2\chi} =\sigma^2 / (8\pi m_\phi)$. For $\Gamma_{\phi\to 2\chi} \lesssim H_0$, model B suffers from the same overclosure problem for $m_\phi\gtrsim 10^{-13}$ eV. 
 If $\Gamma_{\phi\to 2\chi} \simeq H_{\mathrm{BBN}}$, $\phi$ would be subject to BBN constraint, which imposes a condition on model parameters. For $m_\phi \gtrsim \sigma \gtrsim 5\times 10^{-8}\sqrt{m_\phi \times {\rm {eV}}}$ the model satisfies both the perturbative unitarity and the condition that $\phi$ decays before BBN, which we elaborate in the Supplemental Material~\cite{cite:sup}. However, increasing $\sigma$ beyond a certain value can lead to suppression of parametric resonance, making it challenging to find a viable parameter set. We leave such a dedicated parameter scan for future work.

All four models we consider yield new relativistic degrees of freedom beyond the SM, i.e., dark radiation, including the GWs and/or massless scalar(s), potentially contributing to $\Delta N_{\rm {eff}}$ relevant to BBN prediction and CMB observations.
The current limit of $|\Delta N_{\rm{eff}}| \lesssim 0.29$ at 95\% C.L.~\cite{Planck:2018vyg,Pagano:2015hma} is set by the {\it Planck} satellite. Next-generation CMB and large-scale structure experiments (CMB-S4~\cite{Abazajian:2019eic}, COrE~\cite{COrE:2011bfs}, Euclid~\cite{EUCLID:2011zbd}) will be able to probe $\Delta N_{\rm{eff}} ={\cal O}(0.01)$. Our models can yield $\Delta N_{\rm{eff}}$ up to ${\cal O}(0.1)$, while being consistent with the current bound, making these simple models relevant for both the upcoming GW and CMB experiments, offering complementary signals (see Supplemental Material for details on estimating $\Delta N_{\rm {eff}}$~\cite{cite:sup}). 

\par

\begin{table}
	\begin{tabular}{|c|c|c|c|c|c|c|c|}
		\hline	
	Model &	$m_\phi\, (\rm {eV})$            &  $g$     &  $\sigma\, (\rm {eV})$ 
  & $\lambda_\chi$ &  $\nu_{\rm {GW}}$(Hz)          & $\Omega_{\rm {GW}}$                     \\ \hline\hline
		A         & $10^{-13}$       &   $10^{-75}$         &  -          & -    &   $10^{-9}$               & $10^{-10}$            
			\\
			\hline
   A$^*$         & $10^{8}$      &     $10^{-36}$         &  -          & -    &    $100$             & $10^{-9}$              
			\\
			\hline
B   & $10^{-13} $         &  -          &  $10^{-52}$       &  $10^{-75} $      & $10^{-9}$                        & $10^{-9} $                  
 \\
			\hline
 B$^*$   & $10^{-2} $         &  -          &  $10^{-30}$       &  $10^{-53} $      & $10^{-3}$                        & $10^{-9} $                 
 \\
			\hline
 B$^*$  & $10^{8} $          &  -          &  $10^{-10}$       &  $10^{-33} $      & $10^{2}$                        & $10^{-9}$                   
			\\
			\hline\hline
   &	$\lambda_\phi$          & $g$     &  $\sigma\, (\rm {eV})$ 
  & $\lambda_\chi$ &  $\nu_{\rm {GW}}$(Hz)          & $\Omega_{\rm {GW}}$           \\
\hline
C   & $10^{-35}$          &  -          &  -       &  -     & $100$                        &  $10^{-11.5}$                 
\\
\hline
D   & $10^{-79}$           &  $10^{-79}$      &  -       &  -     & $10^{-9}$                        &  $10^{-12}$                                
  \\ \hline
	\end{tabular}
	\caption{Example model parameters leading to GW production with frequencies in observationally relevant range, from NANOGrav to LISA to LIGO. The values of $\Omega_{\rm {GW}}$ indicate the peak amplitude. The energy density of the scalar field is assumed to be $10\%$ of the total energy density of the Universe at the start of the simulation (solid lines in Fig.~\ref{fig:omega_freq}). $\Omega_{\rm {GW}}$ quadratically dependents on the ratio $\rho_\phi /\rho_{\rm {tot}}$. Finally, the points denoted with a $^*$
	lead to DM overproduction for stable $\phi$, but can be viable with additional couplings that allow $\phi$ to decay (see main text for details).
}
	\label{tab1}
\end{table}
\par
\noindent \textit{Discussion and Conclusion.}
The presence of oscillating scalar fields in the early Universe naturally arises in many well-motivated particle physics theories beyond the SM, with the potential of sourcing SGWBs through parametric resonance effects. In this Letter, we conducted a general study of this scenario with a focus on simple representative models and surveyed a wide range of mass and coupling parameters using lattice simulations. Depending on model specifics, we found that such models can give rise to GW signal with amplitude up to $\Omega_{\rm {GW}}\sim10^{-9}$, with frequency ranging from the band of PTA experiments to that of LISA and LIGO/ET/CT~\cite{Janssen:2014dka,LISA:2017pwj,Robson:2018ifk,Crowder:2005nr} and beyond~\cite{Caprini:2019pxz,Sesana:2019vho,Schmitz:2020syl,Garcia-Bellido:2021zgu,Braglia:2021fxn,NANOGrav:2023gor}, and with a distinct, multipeak spectrum in some cases. Intriguing complementary phenomenology is also identified: in the models with a massive scalar $\phi$, the residual relic $\phi$ field can naturally serve as DM candidate, and in particular $m_\phi\sim10^{-13}$ eV marks a viable DM mass range that correlates with GW signal covered by PTA experiments; in the models with massless scalars, the relic scalar radiation contributes to CMB $N_{\rm{eff}}$ observable (in addition to the generally subleading contribution from GWs). Further explorations of these scenarios may lead to even richer findings. 
For instance, while we focused on considering the standard cosmic history of radiation domination during the $\phi$ oscillation period, early matter domination driven by the $\phi$ condensate may amplify the GW signal and affect DM structure formation and DM detection today \cite{Erickcek:2015bda, Erickcek:2015jza}. 
In addition, it is worth further investigating the natural connection with ultralight DM as emerging in some benchmarks and its ramifications, as well as the potential complementarity with CMB observations\cite{Note2}

\medskip

We thank Anish Ghoshal for collaboration during the early stage of this project.
We also thank Soubhik Kumar and Raman Sundrum for helpful discussions.
YC is supported by the US Department of Energy under award number DE-SC0008541.
EIS is supported by a fellowship from ``la Caixa'' Foundation (ID 100010434) and from the European Union's Horizon 2020 research
and innovation programme under the Marie Sk\l{}odowska-Curie grant agreement No 847648, the fellowship code is
LCF/BQ/PI20/11760021. %
PS is supported by Grant-in-Aid for Scientific Research (B) under Contract No.~23H01177 and was supported by Seoul National University of Science and Technology during the earlier stage of this work. Numerical computation in this work was carried out at the 
Yukawa Institute Computer Facility. 
Some earlier simulations were performed at the \texttt{Dark} cluster at Seoultech.


\providecommand{\noopsort}[1]{}\providecommand{\singleletter}[1]{#1}%
\providecommand{\href}[2]{#2}
\raggedright

\medskip

%


\begin{thebibliography}{10}
\bibitem{LIGOScientific:2016aoc}
{\bfseries LIGO Scientific, Virgo} Collaboration, B.~P. Abbott {\em et~al.},
  ``{Observation of Gravitational Waves from a Binary Black Hole Merger},''
  \href{http://dx.doi.org/10.1103/PhysRevLett.116.061102}{{\em Phys. Rev.
  Lett.} {\bfseries 116} no.~6, (2016) 061102},
  \href{http://arxiv.org/abs/1602.03837}{{\ttfamily arXiv:1602.03837 [gr-qc]}}.

\bibitem{NANOGrav:2023gor}
{\bfseries NANOGrav} Collaboration, G.~Agazie {\em et~al.}, ``{The NANOGrav 15
  yr Data Set: Evidence for a Gravitational-wave Background},''
  \href{http://dx.doi.org/10.3847/2041-8213/acdac6}{{\em Astrophys. J. Lett.}
  {\bfseries 951} no.~1, (2023) L8},
  \href{http://arxiv.org/abs/2306.16213}{{\ttfamily arXiv:2306.16213
  [astro-ph.HE]}}.

\bibitem{EPTA:2023fyk}
{\bfseries EPTA} Collaboration, J.~Antoniadis {\em et~al.}, ``{The second data
  release from the European Pulsar Timing Array III. Search for gravitational
  wave signals},'' \href{http://dx.doi.org/10.1051/0004-6361/202346844}{{\em
  Astron. Astrophys.} {\bfseries 678} (2023) A50},
  \href{http://arxiv.org/abs/2306.16214}{{\ttfamily arXiv:2306.16214
  [astro-ph.HE]}}.

\bibitem{EPTA:2023xxk}
{\bfseries EPTA} Collaboration, J.~Antoniadis {\em et~al.}, ``{The second data
  release from the European Pulsar Timing Array: V. Implications for massive
  black holes, dark matter and the early Universe},''
  \href{http://arxiv.org/abs/2306.16227}{{\ttfamily arXiv:2306.16227
  [astro-ph.CO]}}.

\bibitem{Reardon:2023gzh}
D.~J. Reardon {\em et~al.}, ``{Search for an Isotropic Gravitational-wave
  Background with the Parkes Pulsar Timing Array},''
  \href{http://dx.doi.org/10.3847/2041-8213/acdd02}{{\em Astrophys. J. Lett.}
  {\bfseries 951} no.~1, (2023) L6},
  \href{http://arxiv.org/abs/2306.16215}{{\ttfamily arXiv:2306.16215
  [astro-ph.HE]}}.

\bibitem{Xu:2023wog}
H.~Xu {\em et~al.}, ``{Searching for the Nano-Hertz Stochastic Gravitational
  Wave Background with the Chinese Pulsar Timing Array Data Release I},''
  \href{http://dx.doi.org/10.1088/1674-4527/acdfa5}{{\em Res. Astron.
  Astrophys.} {\bfseries 23} no.~7, (2023) 075024},
  \href{http://arxiv.org/abs/2306.16216}{{\ttfamily arXiv:2306.16216
  [astro-ph.HE]}}.

\bibitem{NANOGrav:2023hvm}
{\bfseries NANOGrav} Collaboration, A.~Afzal {\em et~al.}, ``{The NANOGrav 15
  yr Data Set: Search for Signals from New Physics},''
  \href{http://dx.doi.org/10.3847/2041-8213/acdc91}{{\em Astrophys. J. Lett.}
  {\bfseries 951} no.~1, (2023) L11},
  \href{http://arxiv.org/abs/2306.16219}{{\ttfamily arXiv:2306.16219
  [astro-ph.HE]}}.

\bibitem{Boyle:2005se}
L.~A. Boyle and P.~J. Steinhardt, ``{Probing the early universe with
  inflationary gravitational waves},''
  \href{http://dx.doi.org/10.1103/PhysRevD.77.063504}{{\em Phys. Rev. D}
  {\bfseries 77} (2008) 063504},
  \href{http://arxiv.org/abs/astro-ph/0512014}{{\ttfamily
  arXiv:astro-ph/0512014}}.

\bibitem{Boyle:2007zx}
L.~A. Boyle and A.~Buonanno, ``{Relating gravitational wave constraints from
  primordial nucleosynthesis, pulsar timing, laser interferometers, and the
  CMB: Implications for the early Universe},''
  \href{http://dx.doi.org/10.1103/PhysRevD.78.043531}{{\em Phys. Rev. D}
  {\bfseries 78} (2008) 043531},
  \href{http://arxiv.org/abs/0708.2279}{{\ttfamily arXiv:0708.2279
  [astro-ph]}}.

\bibitem{Cui:2017ufi}
Y.~Cui, M.~Lewicki, D.~E. Morrissey, and J.~D. Wells, ``{Cosmic Archaeology
  with Gravitational Waves from Cosmic Strings},''
  \href{http://dx.doi.org/10.1103/PhysRevD.97.123505}{{\em Phys. Rev. D}
  {\bfseries 97} no.~12, (2018) 123505},
  \href{http://arxiv.org/abs/1711.03104}{{\ttfamily arXiv:1711.03104
  [hep-ph]}}.

\bibitem{Cui:2018rwi}
Y.~Cui, M.~Lewicki, D.~E. Morrissey, and J.~D. Wells, ``{Probing the pre-BBN
  universe with gravitational waves from cosmic strings},''
  \href{http://dx.doi.org/10.1007/JHEP01(2019)081}{{\em JHEP} {\bfseries 01}
  (2019) 081}, \href{http://arxiv.org/abs/1808.08968}{{\ttfamily
  arXiv:1808.08968 [hep-ph]}}.

\bibitem{Caldwell:2018giq}
R.~R. Caldwell, T.~L. Smith, and D.~G.~E. Walker, ``{Using a Primordial
  Gravitational Wave Background to Illuminate New Physics},''
  \href{http://dx.doi.org/10.1103/PhysRevD.100.043513}{{\em Phys. Rev. D}
  {\bfseries 100} no.~4, (2019) 043513},
  \href{http://arxiv.org/abs/1812.07577}{{\ttfamily arXiv:1812.07577
  [astro-ph.CO]}}.

\bibitem{Caprini:2018mtu}
C.~Caprini and D.~G. Figueroa, ``{Cosmological Backgrounds of Gravitational
  Waves},'' \href{http://dx.doi.org/10.1088/1361-6382/aac608}{{\em Class.
  Quant. Grav.} {\bfseries 35} no.~16, (2018) 163001},
  \href{http://arxiv.org/abs/1801.04268}{{\ttfamily arXiv:1801.04268
  [astro-ph.CO]}}.

\bibitem{Caldwell:2022qsj}
R.~Caldwell {\em et~al.}, ``{Detection of early-universe gravitational-wave
  signatures and fundamental physics},''
  \href{http://dx.doi.org/10.1007/s10714-022-03027-x}{{\em Gen. Rel. Grav.}
  {\bfseries 54} no.~12, (2022) 156},
  \href{http://arxiv.org/abs/2203.07972}{{\ttfamily arXiv:2203.07972 [gr-qc]}}.

\bibitem{Kitajima:2018zco}
N.~Kitajima, J.~Soda, and Y.~Urakawa, ``{Gravitational wave forest from string
  axiverse},'' \href{http://dx.doi.org/10.1088/1475-7516/2018/10/008}{{\em
  JCAP} {\bfseries 10} (2018) 008},
  \href{http://arxiv.org/abs/1807.07037}{{\ttfamily arXiv:1807.07037
  [astro-ph.CO]}}.

\bibitem{Lozanov:2019ylm}
K.~D. Lozanov and M.~A. Amin, ``{Gravitational perturbations from oscillons and
  transients after inflation},''
  \href{http://dx.doi.org/10.1103/PhysRevD.99.123504}{{\em Phys. Rev. D}
  {\bfseries 99} no.~12, (2019) 123504},
  \href{http://arxiv.org/abs/1902.06736}{{\ttfamily arXiv:1902.06736
  [astro-ph.CO]}}.

\bibitem{Hiramatsu:2020obh}
T.~Hiramatsu, E.~I. Sfakianakis, and M.~Yamaguchi, ``{Gravitational wave
  spectra from oscillon formation after inflation},''
  \href{http://dx.doi.org/10.1007/JHEP03(2021)021}{{\em JHEP} {\bfseries 03}
  (2021) 021}, \href{http://arxiv.org/abs/2011.12201}{{\ttfamily
  arXiv:2011.12201 [hep-ph]}}.

\bibitem{Zhou:2020kkf}
Z.~Zhou, J.~Jiang, Y.-F. Cai, M.~Sasaki, and S.~Pi, ``{Primordial black holes
  and gravitational waves from resonant amplification during inflation},''
  \href{http://dx.doi.org/10.1103/PhysRevD.102.103527}{{\em Phys. Rev. D}
  {\bfseries 102} no.~10, (2020) 103527},
  \href{http://arxiv.org/abs/2010.03537}{{\ttfamily arXiv:2010.03537
  [astro-ph.CO]}}.

\bibitem{Easther:2006vd}
R.~Easther, J.~T. Giblin, Jr., and E.~A. Lim, ``{Gravitational Wave Production
  At The End Of Inflation},''
  \href{http://dx.doi.org/10.1103/PhysRevLett.99.221301}{{\em Phys. Rev. Lett.}
  {\bfseries 99} (2007) 221301},
  \href{http://arxiv.org/abs/astro-ph/0612294}{{\ttfamily
  arXiv:astro-ph/0612294}}.

\bibitem{Easther:2006gt}
R.~Easther and E.~A. Lim, ``{Stochastic gravitational wave production after
  inflation},'' \href{http://dx.doi.org/10.1088/1475-7516/2006/04/010}{{\em
  JCAP} {\bfseries 04} (2006) 010},
  \href{http://arxiv.org/abs/astro-ph/0601617}{{\ttfamily
  arXiv:astro-ph/0601617}}.

\bibitem{Garcia-Bellido:2007nns}
J.~Garcia-Bellido and D.~G. Figueroa, ``{A stochastic background of
  gravitational waves from hybrid preheating},''
  \href{http://dx.doi.org/10.1103/PhysRevLett.98.061302}{{\em Phys. Rev. Lett.}
  {\bfseries 98} (2007) 061302},
  \href{http://arxiv.org/abs/astro-ph/0701014}{{\ttfamily
  arXiv:astro-ph/0701014}}.

\bibitem{Garcia-Bellido:2007fiu}
J.~Garcia-Bellido, D.~G. Figueroa, and A.~Sastre, ``{A Gravitational Wave
  Background from Reheating after Hybrid Inflation},''
  \href{http://dx.doi.org/10.1103/PhysRevD.77.043517}{{\em Phys. Rev. D}
  {\bfseries 77} (2008) 043517},
  \href{http://arxiv.org/abs/0707.0839}{{\ttfamily arXiv:0707.0839 [hep-ph]}}.

\bibitem{Dufaux:2007pt}
J.~F. Dufaux, A.~Bergman, G.~N. Felder, L.~Kofman, and J.-P. Uzan, ``{Theory
  and Numerics of Gravitational Waves from Preheating after Inflation},''
  \href{http://dx.doi.org/10.1103/PhysRevD.76.123517}{{\em Phys. Rev. D}
  {\bfseries 76} (2007) 123517},
  \href{http://arxiv.org/abs/0707.0875}{{\ttfamily arXiv:0707.0875
  [astro-ph]}}.

\bibitem{Easther:2007vj}
R.~Easther, J.~T. Giblin, and E.~A. Lim, ``{Gravitational Waves From the End of
  Inflation: Computational Strategies},''
  \href{http://dx.doi.org/10.1103/PhysRevD.77.103519}{{\em Phys. Rev. D}
  {\bfseries 77} (2008) 103519},
  \href{http://arxiv.org/abs/0712.2991}{{\ttfamily arXiv:0712.2991
  [astro-ph]}}.

\bibitem{Ghoshal:2022jdt}
A.~Ghoshal and P.~Saha, ``{Detectable Gravitational Waves from (P)-reheating
  probes non-thermal Dark Matter},''
  \href{http://arxiv.org/abs/2203.14424}{{\ttfamily arXiv:2203.14424
  [hep-ph]}}.

\bibitem{White:2021hwi}
G.~White, L.~Pearce, D.~Vagie, and A.~Kusenko, ``{Detectable Gravitational Wave
  Signals from Affleck-Dine Baryogenesis},''
  \href{http://dx.doi.org/10.1103/PhysRevLett.127.181601}{{\em Phys. Rev.
  Lett.} {\bfseries 127} no.~18, (2021) 181601},
  \href{http://arxiv.org/abs/2105.11655}{{\ttfamily arXiv:2105.11655
  [hep-ph]}}.

\bibitem{Cui:2021are}
Y.~Cui and E.~I. Sfakianakis, ``{Detectable gravitational wave signals from
  inflationary preheating},''
  \href{http://dx.doi.org/10.1016/j.physletb.2023.137825}{{\em Phys. Lett. B}
  {\bfseries 840} (2023) 137825},
  \href{http://arxiv.org/abs/2112.00762}{{\ttfamily arXiv:2112.00762
  [hep-ph]}}.

\bibitem{Machado:2018nqk}
C.~S. Machado, W.~Ratzinger, P.~Schwaller, and B.~A. Stefanek, ``{Audible
  Axions},'' \href{http://dx.doi.org/10.1007/JHEP01(2019)053}{{\em JHEP}
  {\bfseries 01} (2019) 053}, \href{http://arxiv.org/abs/1811.01950}{{\ttfamily
  arXiv:1811.01950 [hep-ph]}}.

\bibitem{Madge:2021abk}
E.~Madge, W.~Ratzinger, D.~Schmitt, and P.~Schwaller, ``{Audible axions with a
  booster: Stochastic gravitational waves from rotating ALPs},''
  \href{http://dx.doi.org/10.21468/SciPostPhys.12.5.171}{{\em SciPost Phys.}
  {\bfseries 12} no.~5, (2022) 171},
  \href{http://arxiv.org/abs/2111.12730}{{\ttfamily arXiv:2111.12730
  [hep-ph]}}.

\bibitem{Li:2023vuu}
M.~Li, S.~Sun, Q.-S. Yan, and Z.~Zhao, ``{Gravitational waves from axion wave
  production},'' \href{http://arxiv.org/abs/2309.08407}{{\ttfamily
  arXiv:2309.08407 [hep-ph]}}.

\bibitem{Gasparotto:2023psh}
S.~Gasparotto and E.~I. Sfakianakis, ``{Axiverse Birefringence},''
  \href{http://arxiv.org/abs/2306.16355}{{\ttfamily arXiv:2306.16355
  [astro-ph.CO]}}.

\bibitem{cite:sup}
See Supplemental Material [url] for a discussion on initial conditions,
  equations of motion and other simulation details, which includes Refs~\cite{Starobinsky:1994bd,Markkanen:2019kpv,Markkanen:2018gcw,Tenkanen:2019aij,Antusch:2016con,Adshead:2019igv,Amin:2018xfe,Maggiore:1999vm,Adshead:2016xxj,Brust:2013ova,Chacko:2015noa,Dufaux:2006ee}.

\bibitem{Dine:1995kz}
M.~Dine, L.~Randall, and S.~D. Thomas, ``{Baryogenesis from flat directions of
  the supersymmetric standard model},''
  \href{http://dx.doi.org/10.1016/0550-3213(95)00538-2}{{\em Nucl. Phys. B}
  {\bfseries 458} (1996) 291--326},
  \href{http://arxiv.org/abs/hep-ph/9507453}{{\ttfamily arXiv:hep-ph/9507453}}.

\bibitem{Figueroa:2016wxr}
D.~G. Figueroa and F.~Torrenti, ``{Parametric resonance in the early
  Universe\textemdash{}a fitting analysis},''
  \href{http://dx.doi.org/10.1088/1475-7516/2017/02/001}{{\em JCAP} {\bfseries
  02} (2017) 001}, \href{http://arxiv.org/abs/1609.05197}{{\ttfamily
  arXiv:1609.05197 [astro-ph.CO]}}.

\bibitem{Kofman:1997yn}
L.~Kofman, A.~D. Linde, and A.~A. Starobinsky, ``{Towards the theory of
  reheating after inflation},''
  \href{http://dx.doi.org/10.1103/PhysRevD.56.3258}{{\em Phys. Rev. D}
  {\bfseries 56} (1997) 3258--3295},
  \href{http://arxiv.org/abs/hep-ph/9704452}{{\ttfamily arXiv:hep-ph/9704452}}.

\bibitem{Greene:1997fu}
P.~B. Greene, L.~Kofman, A.~D. Linde, and A.~A. Starobinsky, ``{Structure of
  resonance in preheating after inflation},''
  \href{http://dx.doi.org/10.1103/PhysRevD.56.6175}{{\em Phys. Rev. D}
  {\bfseries 56} (1997) 6175--6192},
  \href{http://arxiv.org/abs/hep-ph/9705347}{{\ttfamily arXiv:hep-ph/9705347}}.

\bibitem{Giblin:2014gra}
J.~T. Giblin and E.~Thrane, ``{Estimates of maximum energy density of
  cosmological gravitational-wave backgrounds},''
  \href{http://dx.doi.org/10.1103/PhysRevD.90.107502}{{\em Phys. Rev. D}
  {\bfseries 90} no.~10, (2014) 107502},
  \href{http://arxiv.org/abs/1410.4779}{{\ttfamily arXiv:1410.4779 [gr-qc]}}.

\bibitem{Giblin:2010sp}
J.~T. Giblin, Jr, L.~R. Price, and X.~Siemens, ``{Gravitational Radiation from
  Preheating with Many Fields},''
  \href{http://dx.doi.org/10.1088/1475-7516/2010/08/012}{{\em JCAP} {\bfseries
  08} (2010) 012}, \href{http://arxiv.org/abs/1006.0935}{{\ttfamily
  arXiv:1006.0935 [astro-ph.CO]}}.

\bibitem{Figueroa:2020rrl}
D.~G. Figueroa, A.~Florio, F.~Torrenti, and W.~Valkenburg, ``{The art of
  simulating the early Universe -- Part I},''
  \href{http://dx.doi.org/10.1088/1475-7516/2021/04/035}{{\em JCAP} {\bfseries
  04} (2021) 035}, \href{http://arxiv.org/abs/2006.15122}{{\ttfamily
  arXiv:2006.15122 [astro-ph.CO]}}.

\bibitem{Figueroa:2021yhd}
D.~G. Figueroa, A.~Florio, F.~Torrenti, and W.~Valkenburg, ``{CosmoLattice: A
  modern code for lattice simulations of scalar and gauge field dynamics in an
  expanding universe},''
  \href{http://dx.doi.org/10.1016/j.cpc.2022.108586}{{\em Comput. Phys.
  Commun.} {\bfseries 283} (2023) 108586},
  \href{http://arxiv.org/abs/2102.01031}{{\ttfamily arXiv:2102.01031
  [astro-ph.CO]}}.

\bibitem{Note1}
Similar result of frequency independent $\Omega _{\protect \mathrm {GW}}$ was
  also found in earlier literature, e.g. in the context of preheating with
  scalar fields \cite {Easther:2006vd}.

\bibitem{Agrawal:2016ubh}
P.~Agrawal and R.~Sundrum, ``{Small Vacuum Energy from Small Equivalence
  Violation in Scalar Gravity},''
  \href{http://dx.doi.org/10.1007/JHEP05(2017)144}{{\em JHEP} {\bfseries 05}
  (2017) 144}, \href{http://arxiv.org/abs/1611.07021}{{\ttfamily
  arXiv:1611.07021 [hep-th]}}.

\bibitem{Planck:2018vyg}
{\bfseries Planck} Collaboration, N.~Aghanim {\em et~al.}, ``{Planck 2018
  results. VI. Cosmological parameters},''
  \href{http://dx.doi.org/10.1051/0004-6361/201833910}{{\em Astron. Astrophys.}
  {\bfseries 641} (2020) A6}, \href{http://arxiv.org/abs/1807.06209}{{\ttfamily
  arXiv:1807.06209 [astro-ph.CO]}}. [Erratum: Astron.Astrophys. 652, C4
  (2021)].

\bibitem{Pagano:2015hma}
L.~Pagano, L.~Salvati, and A.~Melchiorri, ``{New constraints on primordial
  gravitational waves from Planck 2015},''
  \href{http://dx.doi.org/10.1016/j.physletb.2016.07.078}{{\em Phys. Lett. B}
  {\bfseries 760} (2016) 823--825},
  \href{http://arxiv.org/abs/1508.02393}{{\ttfamily arXiv:1508.02393
  [astro-ph.CO]}}.

\bibitem{Abazajian:2019eic}
K.~Abazajian {\em et~al.}, ``{CMB-S4 Science Case, Reference Design, and
  Project Plan},'' \href{http://arxiv.org/abs/1907.04473}{{\ttfamily
  arXiv:1907.04473 [astro-ph.IM]}}.

\bibitem{COrE:2011bfs}
{\bfseries COrE} Collaboration, F.~R. Bouchet {\em et~al.}, ``{COrE (Cosmic
  Origins Explorer) A White Paper},''
  \href{http://arxiv.org/abs/1102.2181}{{\ttfamily arXiv:1102.2181
  [astro-ph.CO]}}.

\bibitem{EUCLID:2011zbd}
{\bfseries EUCLID} Collaboration, R.~Laureijs {\em et~al.}, ``{Euclid
  Definition Study Report},'' \href{http://arxiv.org/abs/1110.3193}{{\ttfamily
  arXiv:1110.3193 [astro-ph.CO]}}.

\bibitem{Janssen:2014dka}
G.~Janssen {\em et~al.}, ``{Gravitational wave astronomy with the SKA},''
  \href{http://dx.doi.org/10.22323/1.215.0037}{{\em PoS} {\bfseries AASKA14}
  (2015) 037}, \href{http://arxiv.org/abs/1501.00127}{{\ttfamily
  arXiv:1501.00127 [astro-ph.IM]}}.

\bibitem{LISA:2017pwj}
{\bfseries LISA} Collaboration, P.~Amaro-Seoane {\em et~al.}, ``{Laser
  Interferometer Space Antenna},''
  \href{http://arxiv.org/abs/1702.00786}{{\ttfamily arXiv:1702.00786
  [astro-ph.IM]}}.

\bibitem{Robson:2018ifk}
T.~Robson, N.~J. Cornish, and C.~Liu, ``{The construction and use of LISA
  sensitivity curves},'' \href{http://dx.doi.org/10.1088/1361-6382/ab1101}{{\em
  Class. Quant. Grav.} {\bfseries 36} no.~10, (2019) 105011},
  \href{http://arxiv.org/abs/1803.01944}{{\ttfamily arXiv:1803.01944
  [astro-ph.HE]}}.

\bibitem{Crowder:2005nr}
J.~Crowder and N.~J. Cornish, ``{Beyond LISA: Exploring future gravitational
  wave missions},'' \href{http://dx.doi.org/10.1103/PhysRevD.72.083005}{{\em
  Phys. Rev. D} {\bfseries 72} (2005) 083005},
  \href{http://arxiv.org/abs/gr-qc/0506015}{{\ttfamily arXiv:gr-qc/0506015}}.

\bibitem{Caprini:2019pxz}
C.~Caprini, D.~G. Figueroa, R.~Flauger, G.~Nardini, M.~Peloso, M.~Pieroni,
  A.~Ricciardone, and G.~Tasinato, ``{Reconstructing the spectral shape of a
  stochastic gravitational wave background with LISA},''
  \href{http://dx.doi.org/10.1088/1475-7516/2019/11/017}{{\em JCAP} {\bfseries
  11} (2019) 017}, \href{http://arxiv.org/abs/1906.09244}{{\ttfamily
  arXiv:1906.09244 [astro-ph.CO]}}.

\bibitem{Sesana:2019vho}
A.~Sesana {\em et~al.}, ``{Unveiling the gravitational universe at $\mu$-Hz
  frequencies},'' \href{http://dx.doi.org/10.1007/s10686-021-09709-9}{{\em
  Exper. Astron.} {\bfseries 51} no.~3, (2021) 1333--1383},
  \href{http://arxiv.org/abs/1908.11391}{{\ttfamily arXiv:1908.11391
  [astro-ph.IM]}}.

\bibitem{Schmitz:2020syl}
K.~Schmitz, ``{New Sensitivity Curves for Gravitational-Wave Signals from
  Cosmological Phase Transitions},''
  \href{http://dx.doi.org/10.1007/JHEP01(2021)097}{{\em JHEP} {\bfseries 01}
  (2021) 097}, \href{http://arxiv.org/abs/2002.04615}{{\ttfamily
  arXiv:2002.04615 [hep-ph]}}.

\bibitem{Garcia-Bellido:2021zgu}
J.~Garcia-Bellido, H.~Murayama, and G.~White, ``{Exploring the early Universe
  with Gaia and Theia},''
  \href{http://dx.doi.org/10.1088/1475-7516/2021/12/023}{{\em JCAP} {\bfseries
  12} no.~12, (2021) 023}, \href{http://arxiv.org/abs/2104.04778}{{\ttfamily
  arXiv:2104.04778 [hep-ph]}}.

\bibitem{Braglia:2021fxn}
M.~Braglia and S.~Kuroyanagi, ``{Probing prerecombination physics by the
  cross-correlation of stochastic gravitational waves and CMB anisotropies},''
  \href{http://dx.doi.org/10.1103/PhysRevD.104.123547}{{\em Phys. Rev. D}
  {\bfseries 104} no.~12, (2021) 123547},
  \href{http://arxiv.org/abs/2106.03786}{{\ttfamily arXiv:2106.03786
  [astro-ph.CO]}}.

\bibitem{Erickcek:2015bda}
A.~L. Erickcek, K.~Sinha, and S.~Watson, ``{Bringing Isolated Dark Matter Out
  of Isolation: Late-time Reheating and Indirect Detection},''
  \href{http://dx.doi.org/10.1103/PhysRevD.94.063502}{{\em Phys. Rev. D}
  {\bfseries 94} no.~6, (2016) 063502},
  \href{http://arxiv.org/abs/1510.04291}{{\ttfamily arXiv:1510.04291
  [hep-ph]}}.

\bibitem{Erickcek:2015jza}
A.~L. Erickcek, ``{The Dark Matter Annihilation Boost from Low-Temperature
  Reheating},'' \href{http://dx.doi.org/10.1103/PhysRevD.92.103505}{{\em Phys.
  Rev. D} {\bfseries 92} no.~10, (2015) 103505},
  \href{http://arxiv.org/abs/1504.03335}{{\ttfamily arXiv:1504.03335
  [astro-ph.CO]}}.

\bibitem{Note2}
Also see\cite {Ramberg:2022irf} for a related recent work on potential
  complementarity between GW and CMB observables.

\bibitem{Starobinsky:1994bd}
A.~A. Starobinsky and J.~Yokoyama, ``{Equilibrium state of a selfinteracting
  scalar field in the De Sitter background},''
  \href{http://dx.doi.org/10.1103/PhysRevD.50.6357}{{\em Phys. Rev. D}
  {\bfseries 50} (1994) 6357--6368},
  \href{http://arxiv.org/abs/astro-ph/9407016}{{\ttfamily
  arXiv:astro-ph/9407016}}.

\bibitem{Markkanen:2019kpv}
T.~Markkanen, A.~Rajantie, S.~Stopyra, and T.~Tenkanen, ``{Scalar correlation
  functions in de Sitter space from the stochastic spectral expansion},''
  \href{http://dx.doi.org/10.1088/1475-7516/2019/08/001}{{\em JCAP} {\bfseries
  08} (2019) 001}, \href{http://arxiv.org/abs/1904.11917}{{\ttfamily
  arXiv:1904.11917 [gr-qc]}}.

\bibitem{Markkanen:2018gcw}
T.~Markkanen, A.~Rajantie, and T.~Tenkanen, ``{Spectator Dark Matter},''
  \href{http://dx.doi.org/10.1103/PhysRevD.98.123532}{{\em Phys. Rev. D}
  {\bfseries 98} no.~12, (2018) 123532},
  \href{http://arxiv.org/abs/1811.02586}{{\ttfamily arXiv:1811.02586
  [astro-ph.CO]}}.

 \bibitem{Tenkanen:2019aij}
T.~Tenkanen, ``{Dark matter from scalar field fluctuations},''
\href{http://dx.doi.org/10.1103/PhysRevLett.123.061302}{{\em Phys. Rev. Lett.}
{\bfseries 123}, no.6 (2019) 061302},
\href{http://arxiv.org/abs/1905.01214}{{\ttfamily arXiv:1905.01214 [astro-ph.CO]}}.  

\bibitem{Antusch:2016con}
S.~Antusch, F.~Cefala, and S.~Orani, ``{Gravitational waves from oscillons
  after inflation},''
  \href{http://dx.doi.org/10.1103/PhysRevLett.118.011303}{{\em Phys. Rev.
  Lett.} {\bfseries 118} no.~1, (2017) 011303},
  \href{http://arxiv.org/abs/1607.01314}{{\ttfamily arXiv:1607.01314
  [astro-ph.CO]}}. [Erratum: Phys.Rev.Lett. 120, 219901 (2018)].

\bibitem{Adshead:2019igv}
P.~Adshead, J.~T. Giblin, M.~Pieroni, and Z.~J. Weiner, ``{Constraining Axion
  Inflation with Gravitational Waves across 29 Decades in Frequency},''
  \href{http://dx.doi.org/10.1103/PhysRevLett.124.171301}{{\em Phys. Rev.
  Lett.} {\bfseries 124} no.~17, (2020) 171301},
  \href{http://arxiv.org/abs/1909.12843}{{\ttfamily arXiv:1909.12843
  [astro-ph.CO]}}.

\bibitem{Amin:2018xfe}
M.~A. Amin, J.~Braden, E.~J. Copeland, J.~T. Giblin, C.~Solorio, Z.~J. Weiner,
  and S.-Y. Zhou, ``{Gravitational waves from asymmetric oscillon dynamics?},''
  \href{http://dx.doi.org/10.1103/PhysRevD.98.024040}{{\em Phys. Rev. D}
  {\bfseries 98} (2018) 024040},
  \href{http://arxiv.org/abs/1803.08047}{{\ttfamily arXiv:1803.08047
  [astro-ph.CO]}}.

\bibitem{Maggiore:1999vm}
M.~Maggiore, ``{Gravitational wave experiments and early universe cosmology},''
  \href{http://dx.doi.org/10.1016/S0370-1573(99)00102-7}{{\em Phys. Rept.}
  {\bfseries 331} (2000) 283--367},
  \href{http://arxiv.org/abs/gr-qc/9909001}{{\ttfamily arXiv:gr-qc/9909001}}.

\bibitem{Adshead:2016xxj}
P.~Adshead, Y.~Cui, and J.~Shelton, ``{Chilly Dark Sectors and Asymmetric
  Reheating},'' \href{http://dx.doi.org/10.1007/JHEP06(2016)016}{{\em JHEP}
  {\bfseries 06} (2016) 016}, \href{http://arxiv.org/abs/1604.02458}{{\ttfamily
  arXiv:1604.02458 [hep-ph]}}.

\bibitem{Brust:2013ova}
C.~Brust, D.~E. Kaplan, and M.~T. Walters, ``{New Light Species and the CMB},''
  \href{http://dx.doi.org/10.1007/JHEP12(2013)058}{{\em JHEP} {\bfseries 12}
  (2013) 058}, \href{http://arxiv.org/abs/1303.5379}{{\ttfamily arXiv:1303.5379
  [hep-ph]}}.

\bibitem{Chacko:2015noa}
Z.~Chacko, Y.~Cui, S.~Hong, and T.~Okui, ``{Hidden dark matter sector, dark
  radiation, and the CMB},''
  \href{http://dx.doi.org/10.1103/PhysRevD.92.055033}{{\em Phys. Rev. D}
  {\bfseries 92} (2015) 055033},
  \href{http://arxiv.org/abs/1505.04192}{{\ttfamily arXiv:1505.04192
  [hep-ph]}}.

\bibitem{Dufaux:2006ee}
J.~F. Dufaux, G.~N. Felder, L.~Kofman, M.~Peloso, and D.~Podolsky,
  ``{Preheating with trilinear interactions: Tachyonic resonance},''
  \href{http://dx.doi.org/10.1088/1475-7516/2006/07/006}{{\em JCAP} {\bfseries
  07} (2006) 006}, \href{http://arxiv.org/abs/hep-ph/0602144}{{\ttfamily
  arXiv:hep-ph/0602144}}.

\bibitem{Ramberg:2022irf}
N.~Ramberg, W.~Ratzinger, and P.~Schwaller, ``{One \ensuremath{\mu} to rule
  them all: CMB spectral distortions can probe domain walls, cosmic strings and
  low scale phase transitions},''
  \href{http://dx.doi.org/10.1088/1475-7516/2023/02/039}{{\em JCAP} {\bfseries
  02} (2023) 039}, \href{http://arxiv.org/abs/2209.14313}{{\ttfamily
  arXiv:2209.14313 [hep-ph]}}.

\end{thebibliography}

\begin{thebibliography}{10}

\bibitem{Starobinsky:1994bd}
A.~A. Starobinsky and J.~Yokoyama, ``{Equilibrium state of a selfinteracting
  scalar field in the De Sitter background},''
  \href{http://dx.doi.org/10.1103/PhysRevD.50.6357}{{\em Phys. Rev. D}
  {\bfseries 50} (1994) 6357--6368},
  \href{http://arxiv.org/abs/astro-ph/9407016}{{\ttfamily
  arXiv:astro-ph/9407016}}.

\bibitem{Markkanen:2019kpv}
T.~Markkanen, A.~Rajantie, S.~Stopyra, and T.~Tenkanen, ``{Scalar correlation
  functions in de Sitter space from the stochastic spectral expansion},''
  \href{http://dx.doi.org/10.1088/1475-7516/2019/08/001}{{\em JCAP} {\bfseries
  08} (2019) 001}, \href{http://arxiv.org/abs/1904.11917}{{\ttfamily
  arXiv:1904.11917 [gr-qc]}}.

\bibitem{Markkanen:2018gcw}
T.~Markkanen, A.~Rajantie, and T.~Tenkanen, ``{Spectator Dark Matter},''
  \href{http://dx.doi.org/10.1103/PhysRevD.98.123532}{{\em Phys. Rev. D}
  {\bfseries 98} no.~12, (2018) 123532},
  \href{http://arxiv.org/abs/1811.02586}{{\ttfamily arXiv:1811.02586
  [astro-ph.CO]}}.

\bibitem{Tenkanen:2019aij}
T.~Tenkanen, ``{Dark matter from scalar field fluctuations},''
\href{http://dx.doi.org/10.1103/PhysRevLett.123.061302}{{\em Phys. Rev. Lett.}
{\bfseries 123}, no.6 (2019) 061302},
\href{http://arxiv.org/abs/1905.01214}{{\ttfamily arXiv:1905.01214 [astro-ph.CO]}}.  

\bibitem{Easther:2006gt}
R.~Easther and E.~A. Lim, ``{Stochastic gravitational wave production after
  inflation},'' \href{http://dx.doi.org/10.1088/1475-7516/2006/04/010}{{\em
  JCAP} {\bfseries 04} (2006) 010},
  \href{http://arxiv.org/abs/astro-ph/0601617}{{\ttfamily
  arXiv:astro-ph/0601617}}.

\bibitem{Easther:2006vd}
R.~Easther, J.~T. Giblin, Jr., and E.~A. Lim, ``{Gravitational Wave Production
  At The End Of Inflation},''
  \href{http://dx.doi.org/10.1103/PhysRevLett.99.221301}{{\em Phys. Rev. Lett.}
  {\bfseries 99} (2007) 221301},
  \href{http://arxiv.org/abs/astro-ph/0612294}{{\ttfamily
  arXiv:astro-ph/0612294}}.

\bibitem{Easther:2007vj}
R.~Easther, J.~T. Giblin, and E.~A. Lim, ``{Gravitational Waves From the End of
  Inflation: Computational Strategies},''
  \href{http://dx.doi.org/10.1103/PhysRevD.77.103519}{{\em Phys. Rev. D}
  {\bfseries 77} (2008) 103519},
  \href{http://arxiv.org/abs/0712.2991}{{\ttfamily arXiv:0712.2991
  [astro-ph]}}.

\bibitem{Dufaux:2007pt}
J.~F. Dufaux, A.~Bergman, G.~N. Felder, L.~Kofman, and J.-P. Uzan, ``{Theory
  and Numerics of Gravitational Waves from Preheating after Inflation},''
  \href{http://dx.doi.org/10.1103/PhysRevD.76.123517}{{\em Phys. Rev. D}
  {\bfseries 76} (2007) 123517},
  \href{http://arxiv.org/abs/0707.0875}{{\ttfamily arXiv:0707.0875
  [astro-ph]}}.

\bibitem{Antusch:2016con}
S.~Antusch, F.~Cefala, and S.~Orani, ``{Gravitational waves from oscillons
  after inflation},''
  \href{http://dx.doi.org/10.1103/PhysRevLett.118.011303}{{\em Phys. Rev.
  Lett.} {\bfseries 118} no.~1, (2017) 011303},
  \href{http://arxiv.org/abs/1607.01314}{{\ttfamily arXiv:1607.01314
  [astro-ph.CO]}}. [Erratum: Phys.Rev.Lett. 120, 219901 (2018)].

\bibitem{Adshead:2019igv}
P.~Adshead, J.~T. Giblin, M.~Pieroni, and Z.~J. Weiner, ``{Constraining Axion
  Inflation with Gravitational Waves across 29 Decades in Frequency},''
  \href{http://dx.doi.org/10.1103/PhysRevLett.124.171301}{{\em Phys. Rev.
  Lett.} {\bfseries 124} no.~17, (2020) 171301},
  \href{http://arxiv.org/abs/1909.12843}{{\ttfamily arXiv:1909.12843
  [astro-ph.CO]}}.

\bibitem{Amin:2018xfe}
M.~A. Amin, J.~Braden, E.~J. Copeland, J.~T. Giblin, C.~Solorio, Z.~J. Weiner,
  and S.-Y. Zhou, ``{Gravitational waves from asymmetric oscillon dynamics?},''
  \href{http://dx.doi.org/10.1103/PhysRevD.98.024040}{{\em Phys. Rev. D}
  {\bfseries 98} (2018) 024040},
  \href{http://arxiv.org/abs/1803.08047}{{\ttfamily arXiv:1803.08047
  [astro-ph.CO]}}.

\bibitem{Easther2007}
R.~Easther, J.~T. Giblin, Jr., and E.~A. Lim, ``{Gravitational Wave Production
  At The End Of Inflation},''
  \href{http://dx.doi.org/10.1103/PhysRevLett.99.221301}{{\em Phys. Rev. Lett.}
  {\bfseries 99} (2007) 221301},
  \href{http://arxiv.org/abs/astro-ph/0612294}{{\ttfamily
  arXiv:astro-ph/0612294}}.

\bibitem{Planck:2018vyg}
{\bfseries Planck} Collaboration, N.~Aghanim {\em et~al.}, ``{Planck 2018
  results. VI. Cosmological parameters},''
  \href{http://dx.doi.org/10.1051/0004-6361/201833910}{{\em Astron. Astrophys.}
  {\bfseries 641} (2020) A6}, \href{http://arxiv.org/abs/1807.06209}{{\ttfamily
  arXiv:1807.06209 [astro-ph.CO]}}. [Erratum: Astron.Astrophys. 652, C4
  (2021)].

\bibitem{Pagano:2015hma}
L.~Pagano, L.~Salvati, and A.~Melchiorri, ``{New constraints on primordial
  gravitational waves from Planck 2015},''
  \href{http://dx.doi.org/10.1016/j.physletb.2016.07.078}{{\em Phys. Lett. B}
  {\bfseries 760} (2016) 823--825},
  \href{http://arxiv.org/abs/1508.02393}{{\ttfamily arXiv:1508.02393
  [astro-ph.CO]}}.

\bibitem{Abazajian:2019eic}
K.~Abazajian {\em et~al.}, ``{CMB-S4 Science Case, Reference Design, and
  Project Plan},'' \href{http://arxiv.org/abs/1907.04473}{{\ttfamily
  arXiv:1907.04473 [astro-ph.IM]}}.

\bibitem{COrE:2011bfs}
{\bfseries COrE} Collaboration, F.~R. Bouchet {\em et~al.}, ``{COrE (Cosmic
  Origins Explorer) A White Paper},''
  \href{http://arxiv.org/abs/1102.2181}{{\ttfamily arXiv:1102.2181
  [astro-ph.CO]}}.

\bibitem{EUCLID:2011zbd}
{\bfseries EUCLID} Collaboration, R.~Laureijs {\em et~al.}, ``{Euclid
  Definition Study Report},'' \href{http://arxiv.org/abs/1110.3193}{{\ttfamily
  arXiv:1110.3193 [astro-ph.CO]}}.

\bibitem{Caprini:2018mtu}
C.~Caprini and D.~G. Figueroa, ``{Cosmological Backgrounds of Gravitational
  Waves},'' \href{http://dx.doi.org/10.1088/1361-6382/aac608}{{\em Class.
  Quant. Grav.} {\bfseries 35} no.~16, (2018) 163001},
  \href{http://arxiv.org/abs/1801.04268}{{\ttfamily arXiv:1801.04268
  [astro-ph.CO]}}.

\bibitem{Maggiore:1999vm}
M.~Maggiore, ``{Gravitational wave experiments and early universe cosmology},''
  \href{http://dx.doi.org/10.1016/S0370-1573(99)00102-7}{{\em Phys. Rept.}
  {\bfseries 331} (2000) 283--367},
  \href{http://arxiv.org/abs/gr-qc/9909001}{{\ttfamily arXiv:gr-qc/9909001}}.

\bibitem{Adshead:2016xxj}
P.~Adshead, Y.~Cui, and J.~Shelton, ``{Chilly Dark Sectors and Asymmetric
  Reheating},'' \href{http://dx.doi.org/10.1007/JHEP06(2016)016}{{\em JHEP}
  {\bfseries 06} (2016) 016}, \href{http://arxiv.org/abs/1604.02458}{{\ttfamily
  arXiv:1604.02458 [hep-ph]}}.

\bibitem{Brust:2013ova}
C.~Brust, D.~E. Kaplan, and M.~T. Walters, ``{New Light Species and the CMB},''
  \href{http://dx.doi.org/10.1007/JHEP12(2013)058}{{\em JHEP} {\bfseries 12}
  (2013) 058}, \href{http://arxiv.org/abs/1303.5379}{{\ttfamily arXiv:1303.5379
  [hep-ph]}}.

\bibitem{Chacko:2015noa}
Z.~Chacko, Y.~Cui, S.~Hong, and T.~Okui, ``{Hidden dark matter sector, dark
  radiation, and the CMB},''
  \href{http://dx.doi.org/10.1103/PhysRevD.92.055033}{{\em Phys. Rev. D}
  {\bfseries 92} (2015) 055033},
  \href{http://arxiv.org/abs/1505.04192}{{\ttfamily arXiv:1505.04192
  [hep-ph]}}.

\bibitem{Dufaux:2006ee}
J.~F. Dufaux, G.~N. Felder, L.~Kofman, M.~Peloso, and D.~Podolsky,
  ``{Preheating with trilinear interactions: Tachyonic resonance},''
  \href{http://dx.doi.org/10.1088/1475-7516/2006/07/006}{{\em JCAP} {\bfseries
  07} (2006) 006}, \href{http://arxiv.org/abs/hep-ph/0602144}{{\ttfamily
  arXiv:hep-ph/0602144}}.

\end{thebibliography}

\clearpage
\newpage
\onecolumngrid
\setcounter{secnumdepth}{3}
\setcounter{equation}{0}
\setcounter{figure}{0}
\setcounter{table}{0}
\setcounter{page}{1}
\makeatletter
\renewcommand{\theequation}{S\arabic{equation}}
\renewcommand{\thefigure}{S\arabic{figure}}
\renewcommand{\bibnumfmt}[1]{[#1]}
\renewcommand{\citenumfont}[1]{#1}
\pagestyle{plain}

\begin{center}
\Large{\textbf{Gravitational Wave Symphony from Oscillating Spectator Scalar Fields}}\\
\medskip
\textit{Supplemental Material}\\
\medskip
{Yanou Cui,~Pankaj Saha and Evangelos I. Sfakianakis }
\end{center}

\subsection{Inflationary fluctuations and Initial conditions}
\label{app:ICs}

We consider two scalar fields $\phi$ and $\chi$ coupled to each other during inflation but not to the inflaton. As light spectator fields, they acquire a vacuum expectation value through de Sitter fluctuations, as discussed in Ref.~\cite{Starobinsky:1994bd} for a single self-interacting scalar field. 
Consider the  most general potential that encompasses all models A, B, C, and D:
\beq
V = \frac{\lambda_\phi}{4}\phi^4+\frac{\lambda_\chi }{ 4}\chi^4 + \frac{1}{2}m_\phi^2\phi^2+\frac{g}{2} \phi^2\chi^2
+\sigma \phi \chi^2 \, .
\eeq
The equations of motion are
\begin{align}
\begin{split}
\ddot \phi - \nabla^2 \phi + 3H\dot\phi+ \lambda_\phi \phi^3 + m_\phi^2\phi^2+ g \phi\chi^2 + \sigma \chi^2=& 0
\\
\ddot \chi - \nabla^2 \chi + 3H\dot\chi + \lambda_\chi \chi^3 + g \phi^2\chi + 2 \sigma\phi\chi =&0 \, .
\end{split}
\end{align}
We then multiply each equation by $\phi$ and $\chi$, respectively, and take the average values. Following Ref.~\cite{Starobinsky:1994bd} we get
\begin{align}
\begin{split}
{\partial \over\partial N}\langle \phi^2\rangle &= {H^2\over 4\pi^2} -{2\lambda_\phi\over H^2 } \langle \phi^2\rangle^2 - {2g\over 3H^2} \langle \phi^2\rangle\langle \chi^2\rangle - \frac{2 m_\phi^2}{3H^2} \langle \phi^2\rangle
\\
{\partial \over\partial N}\langle \chi^2\rangle & = {H^2\over 4\pi^2} -{2\lambda_\chi\over H^2 } \langle \chi^2\rangle^2 - {2g\over 3H^2} \langle \phi^2\rangle\langle \chi^2\rangle,
\end{split}
\end{align}
where we considered $\langle \phi\rangle=0$ and thus the trilinear coupling $\sigma\phi\chi^2$ does not appear in the equation for the variances. Furthermore, we wrote the equations in terms of the number of $e$-folds $N=Nt$.
Both fields start at $\langle \phi^2\rangle 
=0 = \langle\chi^2\rangle$ and initially start growing as $(H^2/ 4\pi^2 )N$ for $m_\phi \ll H$. We can solve for the late-time (asymptotic) values by setting ${\partial/ \partial N}=0$. We assume $g>0$ for simplicity, and obtain
\begin{align}
\begin{split}
\langle \phi^2 \rangle=
\sqrt\frac{3}{2}
\frac{H^2}{2\pi}
 \sqrt{\frac{ \sqrt{\lambda_\chi}}{3 \lambda_\phi \sqrt{\lambda_\chi}+ g \sqrt{\lambda_\phi}}}
\\
\langle \chi^2 \rangle =
\sqrt\frac{3}{2}
\frac{H^2}{2\pi}
 \sqrt{\frac{ \sqrt{\lambda_\phi}}{3 \lambda_\chi \sqrt{\lambda_\phi}+ g \sqrt{\lambda_\chi}}}
 \end{split}
 \label{eq:vevasymptotic}
\end{align}
If the two fields are decoupled ($g=0$), we recover the familiar single-field result
\beq
\label{eq:starobinskyVEV}
\langle \phi^2 \rangle={H^2\over 2\sqrt{2}\,\pi \sqrt{\lambda_\phi}}
\eeq
In our simulations, we require only one of the fields, $\phi$, to have a non-zero VEV, while the other field, $\chi$, which is present in models A, B, and D, to be at the origin. If both fields are light spectators during inflation, this will not be true. We examined some numerical examples with a non-zero initial VEV for $\chi$, which are found to lead to a suppression of parametric resonance. Therefore, $\langle \phi^2 \rangle \gg \langle \chi^2\rangle$ is required to achieve efficient parametric resonance. As discussed in the main text, this condition may not be ad hoc but rather can be realized naturally realized by coupling the $\chi$ field to inflaton or particles in the thermal bath (including SM states). Such couplings could introduce a large mass term for $\chi$, stabilizing it at the origin during or after inflation but before the $\phi$ field starts rolling and oscillating.

Now, let us consider the dynamics of the $\phi$ field alone. As the $\phi$ mass can be neglected during inflation, the field variance $\langle \phi^2\rangle$ grows linearly with the number of $e$-folds of inflation. A requirement for reaching a VEV of the order of the Planck mass is $N\sim 100 (M_{\rm {Pl}}/H_I)^2 \gtrsim 10^{12}$, where we took $H_I\lesssim 10^{-5} M_{\rm {Pl}}$ as constrained by the CMB data. We thus see that a very large number of $e$-folds are required. If a quartic self-coupling is present, the distribution of $\phi$ reaches an equilibrium of Eq.~\eqref{eq:starobinskyVEV}, whereby for a Planckian field VEV, the self-coupling must be $\lambda_\phi \lesssim 10^{-20}$, depending on the ratio $H_I/M_{\rm {Pl}}$. In the case of a massive light free scalar field during inflation, the equilibrium distribution leads to $\sqrt{\langle \phi^2 \rangle}\sim H^2/m_\phi$, which can also reach Planckian values for $m_\phi \ll H$. As an example, we can consider one of the benchmark points for model B as shown in the main text, with $m_\phi = 10^{-13}\, {\rm {eV}}$. Assuming that $\phi_{\rm in}$ arises solely due to de-Sitter fluctuation, we would be lead to a definite prediction for the Hubble scale of inflation. In particular, in order to produce $\langle \phi^2\rangle \sim M_{\rm {Pl}}^2$ as we found for producing sufficient $\Omega_{\rm GW}$, we found that $H_I\sim 10^{-2}\, {\rm {GeV}}$ is required, along with the necessity of very many $e$-folds of inflation. Nevertheless, as noted, $\phi_{\mathrm{in}}$ may be dominantly determined by an initial condition, without imposing specific requirements on inflation. 

Finally, we must note that the quantum fluctuations that generate $\phi_{\rm {in}}$ are uncorrelated with the density perturbations sourced by the inflaton, leading to isocurvature perturbations if $\phi$ is stable. 
For an initial displacement $\phi_{\rm in}$, the isocurvature fluctuations of the $\phi$ field scale as $P_S \sim (\delta\phi/\phi_{\rm {in}})^2 \sim  (H_I/\phi_{\rm {in}})^2$ (see e.g. Ref.~\cite{Tenkanen:2019aij}). For the desirable value of $\phi_{\rm {in}}\sim M_{{\rm Pl}}$ we get $P_{S}\sim H^2/M_{\rm {Pl}}^2$.
 CMB data constrains the amount of isocurvature perturbations compared to the adiabatic density perturbations as
$
{\cal P}_S\lesssim 0.04 {\cal P}_\zeta/f 
$,
where ${\cal P}_{\zeta} = 2.2\times 10^{-9}$, $f\leq1$ is the fraction of $\phi$ energy in the total dark matter (model A and B) or total radiation energy (model C and D). 
Putting everything together, we found the resultant constraint is $H_I\lesssim 10^{-5} M_{\rm {Pl}}/f$, which imposes a constraints on $H_I$ that is comparable to or weaker than what is already required by tensor to scalar ratio in CMB data. Our benchmark model parameters are fully consistent with this bound.

\subsection{Lattice simulation details} 
\subsubsection{Lattice simulations set-up}
We solve the evolution equations for the two fields
\begin{align}
	\label{eq:phi}
	&\ddot{\phi} + 3H\dot{\phi} - \frac{1}{a^2}|\nabla\phi|^2 + \frac{\partial}{\partial\phi}V(\phi,\chi) =0,\\
	\label{eq:chi}
		&\ddot{\chi} + 3H\dot{\chi} - \frac{1}{a^2}|\nabla\chi|^2 + \frac{\partial}{\partial\chi}V(\phi,\chi) =0
\end{align}
tat each lattice point, with the scale factor for the fixed background expansion
\begin{align}
\label{eq:Hub}
a(t) = a(t_{\ast})\left(1 + \frac{3(1+w)}{2}H(t_{\ast})(t-t_{\ast})\right)^{\frac{2}{3(1+w)}},
 \end{align}
 where $a(t_{\ast})$ and $H(t_{\ast})$ are, respectively, the scale factor and Hubble parameter at the beginning of the simulation at $t=t_{\ast}$. For the case of expansion in the radiation-dominated era, we consider the constant equation of state parameter ($w$) for the external fluid sourcing this expansion to be $w=1/3$. This expression can be readily used to get the Hubble expansion rate for any given background of cosmic evolution, such as radiation or matter domination.
\par
The GWs are computed using  the linearized equation of motion of the transverse-traceless tensor perturbations $h_{ij}$:
\begin{align}
    \ddot{h}_{ij} + 3H\dot{h}_{ij} - \frac{\nabla^2h_{ij}}{a^2} = \frac{2}{\Mp^2 a^2}\Pi_{ij}^{TT},
    \label{eq:eqhij}
\end{align}
where $\Pi_{ij}^{TT}$ represents the transverse-traceless part of the effective anisotropic stress tensor $\Pi_{ij}= \sum_{a}\partial_{i}\phi_a\partial_{j}\phi_a$ for the system of scalar fields $\phi_a, \{a = 1,2,\cdots\}$.
The GW energy density is given by
\begin{equation}
    \rho_{\rm GW}(t) = \frac{\Mp^2}{4}\langle\dot{h}_{ij}(\textbf{x},t)\dot{h}_{ij}(\textbf{x},t)\rangle_{\mathcal{V}},
\end{equation}
where $\langle\cdots\rangle_{\mathcal{V}}$ denotes a spatial average over $\mathcal{V}$.
Since we are in a radiation-dominated Universe, assuming that the expansion of the Universe obeys entropy conservation from the end of simulation (subscript ``$\mathrm{e}$") to the present time (subscript ``$0$"), the spectrum of the energy density of GWs (per logarithmic momentum interval) observable today is~\cite{Easther:2006gt,Easther:2006vd,Easther:2007vj,Dufaux:2007pt}:
\begin{align}
\Omega_{\rm GW,0}h^2 = \frac{h^2}{\rho_{\mathrm{crit}}}\frac{d\rho_{\rm GW}}{d\ln k}\Bigg|_{t=t_0} = \frac{h^2}{\rho_{\mathrm{crit}}}\frac{d\rho_{\rm GW}}{d\ln k}\Bigg|_{t=t_\mathrm{e}}\frac{a_\mathrm{e}^4\rho_{\mathrm{e}}}{a_0^4\rho_{\mathrm{crit},0}}= \Omega_{\mathrm{rad}, 0}h^2\Omega_{\mathrm{GW},\mathrm{e}}\left(\frac{g_{\ast}}{g_0}\right)^{-1/3},
\end{align}
where the critical density of the Universe is $\rho_{\mathrm{crit}} = 3\Mp^2H^2$ and $\Omega_{\mathrm{rad}, 0}h^2 = h^2\rho_{\mathrm{rad},0}/\rho_{\mathrm{crit}} = 4.3\times 10^{-5}$.
\par
 Solving these equations numerically on a lattice requires appropriately rescaling the variables so that the equations are reformulated in terms of dimensionless quantities for numerical stability. We worked with the following rescaling for the field and spacetime variables:
\begin{align}
\phi_{\rm {pr}} = \frac{\phi}{\phi_{\rm in}};\quad\chi_{\rm {pr}} = \frac{\chi}{\phi_{\rm in}};\quad
dt_{\rm {pr}} = a^{-\alpha}m_{\mathrm{eff}}dt;\quad k_{\mathrm{pr}} = k/m_{\mathrm{eff}}
\end{align}
where the subscript ``pr" refers to ``program'' variables. The parameter $\alpha$ is chosen from the exponent of the dominating (classical) scalar component, i.e., for a scalar potential whose leading term around the potential minima is $V(\phi)\propto\phi^n$, we have $\alpha = 3(n-2)/(n+2)$. The effective mass  is defined as:
\begin{align}
m_{\mathrm{eff}} &= m_{\phi}\quad\qquad \text{for models~A and B,}\\
m_{\mathrm{eff}} &= \sqrt{\lambda_{\phi}}\phi_{\mathrm{in}}\quad \text{for models~C and D.}
\end{align}
The tensor perturbations are rescaled in a similar way.
\par
\subsubsection{Dimensionless equations}
We can understand the dynamics of parametric resonance and the scaling between different parameter choices (leading to different GW frequencies) by using properly rescaled equations. We  consider Model A for concreteness, where the equations for the background $\phi$ field and $\delta\chi_k$ modes are
\begin{align}
\frac{\dx^2 \phi / \phi_{\rm {in}} }{ \dx(m_\phi t)^2}
+ 3 \frac{H}{ m_\phi} \frac{\dx\phi/\phi_{\rm {in}} }{\dx(m_\phi t)} - \frac{\nabla^2(\phi/\phi_{\rm {in}})}{m^2_\phi a^2} + 
(\phi/\phi_{\rm {in}}) =0
\\
\frac{\dx^2 \delta \chi_k}{ \dx(m_\phi t)^2} + 3 \frac{H}{m_\phi} \frac{\dx\delta \chi_k }{\dx(m_\phi t)} + \frac{k^2}{a^2m_\phi^2}\delta\chi_{ k}
+ g \frac{\phi^2}{m_\phi^2} \delta \chi_k =0
\, ,
\end{align}
where we divided the equation of motion for $\phi$ by $m_\phi^2 \phi_{\rm {in}}$ and the equation of motion for $\delta\chi_k$ by $m_\phi^2$.
We first see how the resonance parameter $q$ arises in this model. Furthermore, by defining  $\tilde t = m_\phi t$, $\tilde H = H/m_\phi$, $\tilde k = k/m_\phi$ and defining a function $\tilde f(\tilde t)$ which contains the time-dependence of the background, such that $g\phi^2 / m_\phi^2 \equiv q \tilde f(\tilde t)$, we see that the equation of motion for the $\chi$ fluctuations is written as
\beq
{\dx^2 \delta\chi_{\tilde k} \over \dx\tilde t^2} + 3 \tilde H {\dx\delta\chi_{\tilde k}\over \dx\tilde t} + \frac{\tilde{k}^2}{a^2} \delta\chi_{\tilde k}+ q f(\tilde t) \delta\chi_{\tilde k}=0
\label{eq:deltachirescaled}
\eeq
The equation has ``lost'' all information about specific scales thus, the growing modes will behave as
$\delta\chi_{\tilde k} \propto e^{\tilde \mu_{\tilde k} \tilde t}$, where $\tilde \mu$ is the Floquet exponent of Eq.~\eqref{eq:deltachirescaled} and is dimensionless. In physical units $\tilde \mu  \tilde t = (\tilde \mu m_\phi) t$, where the dimensionful physical Floquet exponent is $\mu = \tilde\mu \cdot m_\phi$. Thus, we can say that we measure time, Floquet exponents and wavenumbers in units of $m_\phi$.
\par
The only ``memory'' of the mass-scale $m_\phi$ comes through the initial conditions of the $\chi$ fluctuations, which follow the Bunch-Davies vacuum and thus without amplification from parametric resonance $\rho_{\delta\chi} \sim \int_{0}^{\mathcal{O}(H)}d^3k \, k^2 |\delta \chi|^2 \sim H^4\sim m_\phi^4$. 
The amplification will effectively multiply this result by $e^{2\tilde \mu \tilde t}$ and parametric resonance completes when $\rho_{\delta\chi} \sim \rho_\phi \sim m_\phi^2 M_{\rm {Pl}}^2$ or $\tilde \mu \tilde t \sim \log (M_{\rm {Pl}}/m_\phi)$. Because of this fact, parametric resonance will take longer to transfer the energy from the $\phi$ field to $\chi$ fluctuations for smaller values of $m_\phi$, corresponding to smaller GW frequencies. 
\\
\subsubsection{Parameter choice}
We can demonstrate the various parameter choices with a specific example. Let us fix the dynamics to take place around $H =1 \, {\rm {GeV}}$. In a massive scalar field model (e.g., model A), we choose $m_\phi=1 \, {\rm {GeV}}$. 
   Choosing when the field starts rolling largely determines the frequency of GWs. 
    Since we want the scalar field to have subdominant but significant fraction of the energy density of the universe, say one percent, we can estimate $3H^2M_{\rm {Pl}}^2 = 0.01 m_\phi^2 \phi_{\rm {in}}^2/2$, leading to $\phi_{\rm {in}}={\cal O}(M_{\rm {Pl}})$. Given these numbers, we can check what coupling is required to achieve a resonance parameter of $ q \equiv g \phi_{\rm {in}}^2/m_\phi^2={\cal O}(10^4)$ leading to $g\propto m_\phi^2/\phi_{\rm {in}}^2 \sim (1\, {\rm {GeV}}  / M_{\rm {Pl}})^2 $, which is the reason for the small coupling between the two fields in this simple analysis of model A.
    \\
\subsubsection{Convergence tests}
\begin{figure}[h!]
    \centering
    \includegraphics[scale=0.75]{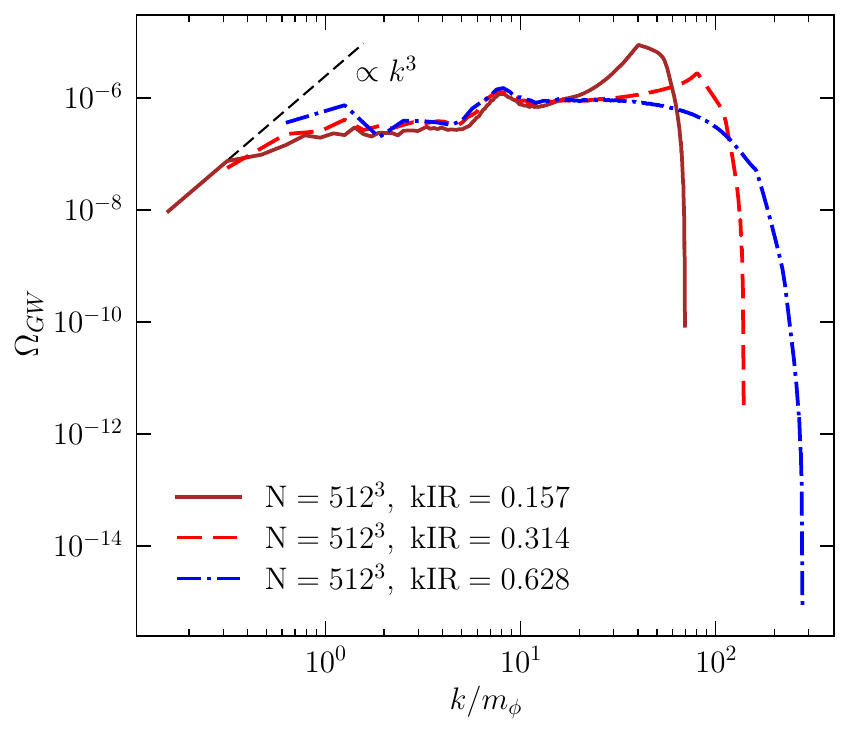}
    \caption{The GW spectrum for Model B for different box sizes, exhibiting convergence in the central region of wave-numbers, allowing us to resolve the $k^3$ scaling at low wave numbers. 
    A noted earlier in the literature~\cite{Easther2007,Antusch:2016con,Amin:2018xfe,Adshead:2019igv}, the UV parts of the spectra for some choices of $k_{\mathrm{IR}}$ showing a growth as $k^4$ are numerical artifacts and do not correspond to any physical result. With an optimal choice of lattice parameters (for instance, the blue dash-dotted line in this figure), such UV ``peaks'' can be avoided.}
    \label{fig:convergence}
\end{figure}

The resolution of the simulation is limited by the comoving box size (L) which sets the IR momenta cutoff of the simulation $k_{\rm IR} = 2\pi/L$, and the number of lattice points $N$ which sets the UV cutoff $k_{\rm UV} = \sqrt{3}/2Nk_{\rm IR}$ (for a $3+1$ dimensional simulation).
The appearance of unphysical peaks at the high frequency end of the spectrum is also expected due to noise towards the unresolved Nyquist frequencies and noise due to finite differencing. Such ``UV peaks" have been well documented in the literature \cite{Easther:2007vj,Antusch:2016con,Adshead:2019igv} and have been studied in detail in Ref.~\cite{Amin:2018xfe}.
On the other hand, to make sure that the modes we are interested in resolving are within the simulation box, we had to choose a sufficiently large box size.
This limits our range of $k_{\rm IR}$. 
It is thus necessary to check the convergence of the simulations by performing multiple simulations with different lattice points and box sizes. 
The results, which are physical, do not change significantly if the box size and resolution are chosen properly.
To avoid unwanted UV artifacts, we performed all simulations with $512^3$ lattice sites with optimal $k_{\rm IR}$ for our main results in Fig. 1 of the main text.
Meanwhile, in each simulation, we made sure that we did not lose any IR peak from the spectra. Similarly, in choosing the lattice points, we have made sure that we do not get any UV artifacts. 
Such a trade-off results in missing the infrared ``tail'' for some spectra. 
That being said, Figure~\ref{fig:convergence} shows three different simulations of model B for the same number of lattice points $512^3$ and different box sizes (equivalently, different $k_{\rm {IR}}$). 
We can see that reducing $k_{\rm {IR}}$ for the same number of grid points reduces $k_{\rm {UV}}$ and at the same time introduces an unphysical peak close to $k_{\rm {UV}}$.
However, in the intermediate wavenumber regime, all three simulations match very well. 
Furthermore, we see that the simulation with the smallest $k_{\rm {IR}}$ does indeed exhibit the expected $k^3$ behavior at low wavenumber.
The GW spectra shown in the main text are computed by using an infrared (IR) cut-off $k_{\mathrm{IR,pr}}\in [0.157,3]$ depending on the model.

\subsection{New relativistic degrees of freedom and contribution to $\Delta N_{\rm{eff}}$}
The models we consider naturally yield new relativistic degrees of freedom beyond the SM or dark radiation, potentially contributing to $\Delta N_{\rm{eff}}$ relevant to BBN prediction and CMB observations.
The current limit of $|\Delta N_{\rm{eff}}| \lesssim 0.29$ at 95\% C.L.~\cite{Planck:2018vyg,Pagano:2015hma} is set by the data collected by {\it Planck} satellite. Next-generation CMB-S4 experiments will be able to probe $\Delta N_{\rm{eff}} \lesssim 0.06$ at $2\sigma$~\cite{Abazajian:2019eic}, whereas the next generation of satellite missions like COrE~\cite{COrE:2011bfs} and Euclid~\cite{EUCLID:2011zbd} will impose bounds at $2\sigma$ on $\Delta N_{\rm{eff}}\lesssim 0.013$.
It is thus important to estimate this effect in order to ensure that the related constraints are satisfied and to uncover a potentially complementary signature of these models. 

For all the benchmark models we consider, GW is a universal source of contribution to $\Delta N_{\rm{eff}}$. This contribution is easily computed as \cite{Caprini:2018mtu,Adshead:2019igv, Maggiore:1999vm}
\beq
{\Omega_{\rm{GW},0}h^2\over \Omega_{\gamma,0}h^2} = {7\over 8}\left ({4\over 11}\right )^{4/3}
\Delta N_{\rm{eff}} \eeq
By using $\Omega_{\gamma,0}h^2 \simeq 2.47\cdot 10^{-5}$ we find that
\beq
\Delta N_{\rm{eff}} \simeq 10^4 \Omega_{\rm{GW},0}
\eeq
Since the produced GW amplitude today in our models is at the level of $\lesssim {\cal O}(10^{-9})$, the contribution to $\Delta N_{\rm{eff}}$ is negligible and well below the sensitivities of foreseeable CMB experiments.

All the models we consider contain at least one massless scalar; $\chi$ for models A and B, $\phi$ for model C and both for model D. This leads to additional contributions to $\Delta N_{\rm{eff}}$ that typically dominate over that from GWs as estimated above. There are two components of such dark radiation: the thermal background component present prior to the onset of the $\phi$ condensates' oscillation, i.e., the relic from the primordial reheating stage, and the non-thermally produced component due to resonance particle production during the $\phi$ condensates' oscillation. The thermal background component depends on reheating details beyond the scope of this paper and can easily be negligibly small compared to the non-thermal component. For instance, the dark sector can be much colder than the SM due to asymmetric reheating and never thermalized with the SM due to feeble interactions \cite{Adshead:2016xxj}. Therefore we choose to focus on the latter, non-thermal component. In order to estimate this contribution, we make a few simple, reasonable assumptions. First, after production, the dark radiation does not thermalize with the SM, which is consistent with our minimal models where non-gravitational coupling to the SM is absent. Second, we assume that  particle production from the oscillation of the $\phi$ condense  completes quickly (i.e., within O(1) Hubble time) after the onset of oscillation at $T_{\rm {osc}}$ 
 (in analogy to ``instant reheating"), and the fraction of energy released in the form of dark radiation is simply parametrized by a constant $\xi$.  This assumption is generally consistent with the efficient parametric resonance effect we consider here for detectable GW signal and enables a simple yet reasonable estimate of the effect. For models C and D, $\xi=100\%$ is easily justified, while for models A and B, $\xi\lesssim 1$ due to the possible presence of residual massive $\phi$, in the form of a condensate or fluctuations. Nevertheless, for models A and B, $\xi=100\%$ can be taken for a conservative estimate for the maximal $\Delta N_{\rm{eff}}$. A more accurate determination of $\Delta N_{\rm{eff}}$ from massless $\chi$ ($\phi$) from $\phi$ condensate requires a dedicated numerical tracking of the evolution of the particles following production, which we will defer for further investigation. The estimate given here is sufficient to provide the rough estimate/upper bounds necessary for our current analysis.

We may now proceed to estimate  $\Delta N_{\rm{eff}}$  following the arguments of  \cite{Brust:2013ova, Chacko:2015noa}. Note that most existing literature considers the scenario where the dark radiation sector was once in thermal equilibrium with the SM, then thermal decoupling occurs at a later time, after which comoving entropy is conserved separately in each sector. The scenario we consider here is very different in that we consider non-thermally produced dark radiation that never equilibrates with the SM and may or may not self-thermalize following the production (depending on the model couplings); thus, entropy conservation may or may not apply well for the dark sector. Instead of tracking the energy or entropy evolution in the two sectors starting from the time of thermal decoupling, here, the proper starting point is the onset of $\phi$ oscillations at $T_{\rm osc}$. For a straightforward application of the procedure of $\Delta N_{\rm{eff}}$ as given in \cite{Chacko:2015noa}, we assume that self-coupling is effective enough to thermalize the dark radiation species among themselves, thus entropy conservation can apply to the dark sector. By applying entropy conservation in the SM and dark sector starting from $T_{\rm {osc}}$, we find:
\beq \label{Eq:S_cons}
\frac{\hat{g}_*\hat{T}^3}{g_*T^3}=\frac{\hat{g}_{*, \rm {osc}}\hat{T}_{\rm osc}^3}{g_{*,\rm {osc}} T_{\rm {osc}}^3} \, ,
\eeq
where the quantities with $\hat{}$ (all in the numerator) are of the dark sector, while those without $\hat{}$ (in the denominator) are 
of the SM; $g_*$, $T$ are evaluated at any time after $T_{\rm {osc}}$. Strictly speaking, the effective number of relativistic degrees of freedom $g_*$ in the above equation should be $g_{*s}$; however, for the estimate we do here, it is the same as $g_*$. Next, we apply the already introduced input parameter $\alpha$, which is the ratio of total energy density stored in $\phi$ condensate over the total cosmic energy density (assumed to be radiation domination) at $T_{\rm {osc}}$, as well as the $\xi$ parameter introduced earlier as the fraction of $\phi$ energy released into dark radiation, and find:
\beq \label{Eq:alpha}
\alpha\equiv\left(\frac{\rho_{\phi}}{\rho_{\rm {tot}}}\right)_{\rm {osc}}=\frac{\hat{g}_{*, \rm {osc}}\hat{T}_{\rm {osc}}^4}{\xi g_{*,\rm {osc}} T_{\rm {osc}}^4}.
\eeq
Using Eqs.~\ref{Eq:S_cons},\ref{Eq:alpha}, and evaluating  just above SM neutrino coupling at $T\sim O(10)$ MeV when $g_*=10.75$, we derive $\Delta N_{\rm{eff}}$ (applicable at the CMB time) as:
\beq \label{Eq:Neff}
\Delta N_{\rm{eff}}=\frac{\hat{g}_*\hat{T}^4}{\frac{7}{4}T^4}=\frac{4}{7}\alpha\xi\hat{g}_*\left(\frac{g_*}{\hat{g}_*}\right)^{4/3}\left(\frac{\hat{g}_{*,\rm {osc}}}{{g}_{*,\rm {osc}}}\right)^{1/3}.
\eeq
Although Eq.~\ref{Eq:Neff} was formally derived assuming entropy conservation in a thermalized dark sector, essentially the same result can be reached even if the dark sector is not thermalized: in the latter case we would instead directly apply the redshift of the energy density of dark radiation and take $\hat{g}_*=\hat{g}_{*,\rm {osc}}=1$ or $2$ depending on the model ($\hat{g}$ and $\hat{g}_{*,\rm {osc}}$ factors then cancel in Eq.~\ref{Eq:Neff}). 

Plugging in  $g_*=10.75$ and the model-dependent $T_{\rm {osc}}$ and ${g}_{*,\rm {osc}}$, we can find the value of $\Delta N_{\rm{eff}}$. For $T_{\rm {osc}}\gtrsim T_{\rm {EW}}\sim 100$ GeV, ${g}_{*,\rm {osc}}=106.75$, 
with $\alpha\sim10\%$, $\xi\sim 1$, we find $\Delta N_{\rm{eff}}\sim O(0.1)$, readily compatible with existing constraint and potentially detectable with near-future experiments. 
With $\alpha\sim O(10)\%$ yet much lower $T_{\rm {osc}}$, $\Delta N_{\rm{eff}}$ could be in conflict with the Planck bound. Nevertheless, keep in mind that for models A and B, $\Delta N_{\rm{eff}}$ as estimated here is a conservative upper bound. For $\alpha\sim O(1)\%$, $\Delta N_{\rm{eff}}\sim O(0.01)$, thus is generally safe from existing bound while being detectable with future observations. In the main text, we have given the estimates of $H_{\rm {osc}}$ for the models we consider, which relates to $T_{\rm {osc}}$ according to $H_{\rm {osc}}^2= (\pi^2 g / 90)T_{\rm {osc}}^4/M_{\rm {Pl}}^2$. Therefore, the result given here can be readily converted to the constraint/region of interest for model parameters such as $m_\phi$ and the couplings that determine $H_{\rm {osc}}$. In the following, we will give a concrete example based on a simplified numerical analysis, which is consistent with the estimates given above.
Consider model B with $m_{\phi} \sim 10^{-13}$eV, leading to a GW frequency in the nHz range. Here $\chi$ represents the dark radiation contributing to $N_{\rm {eff}}$. By numerically tracking our simulation, we find that approximately $60\%$ of the total $\phi$ energy is transferred to $\chi$ at the end of the simulation for typical values of coupling, as shown in Table 1. Hence, we take $\xi\simeq0.6$ and neglect the variation of the effective number of relativistic d.o.f's, and find  $\Delta N_{\mathrm{eff}} = 0.016$ assuming $\alpha=0.01$ initially. 



\subsection{Trilinear coupling, backreaction, and DM}

It is worth studying the evolution of the background field $\phi$ in more detail. The equation of motion for the $\phi$ field is
\beq
\ddot\phi  +3H\dot\phi - {\nabla^2\phi \over a^2} +m^2_\phi\phi + \sigma \chi^2=0
\eeq
When considering only the spatially averaged field value $\phi_{\rm av}$, we drop the gradient term. The treatment of the term $\chi^2$ requires more care. Initially, it is a subleading term when the field $\chi$ starts in its Bunch-Davies vacuum. As $\chi$ grows, at some point, the variance of $\chi$ becomes comparable to $\phi_{\rm av}^2$, and back-reaction becomes important. We can then use the Hartree approximation to write \cite{Dufaux:2006ee}
\beq
\ddot\phi_{\rm {av}}  +3H\dot\phi_{\rm {av}}  +m^2_\phi\phi_{\rm {av}}  + \sigma \langle\chi^2\rangle=0
\eeq
Neglecting the second time derivative and the Hubble friction, we find
\beq
\phi_{\rm {av}}\simeq - {\sigma\over 2m_\phi^2 }\langle\chi^2\rangle \, ,
\eeq
thus $\phi_{\rm {av}}$ follows the evolution of $\langle \chi^2\rangle$.
\begin{figure}
    \centering    \includegraphics[width=0.5\textwidth]{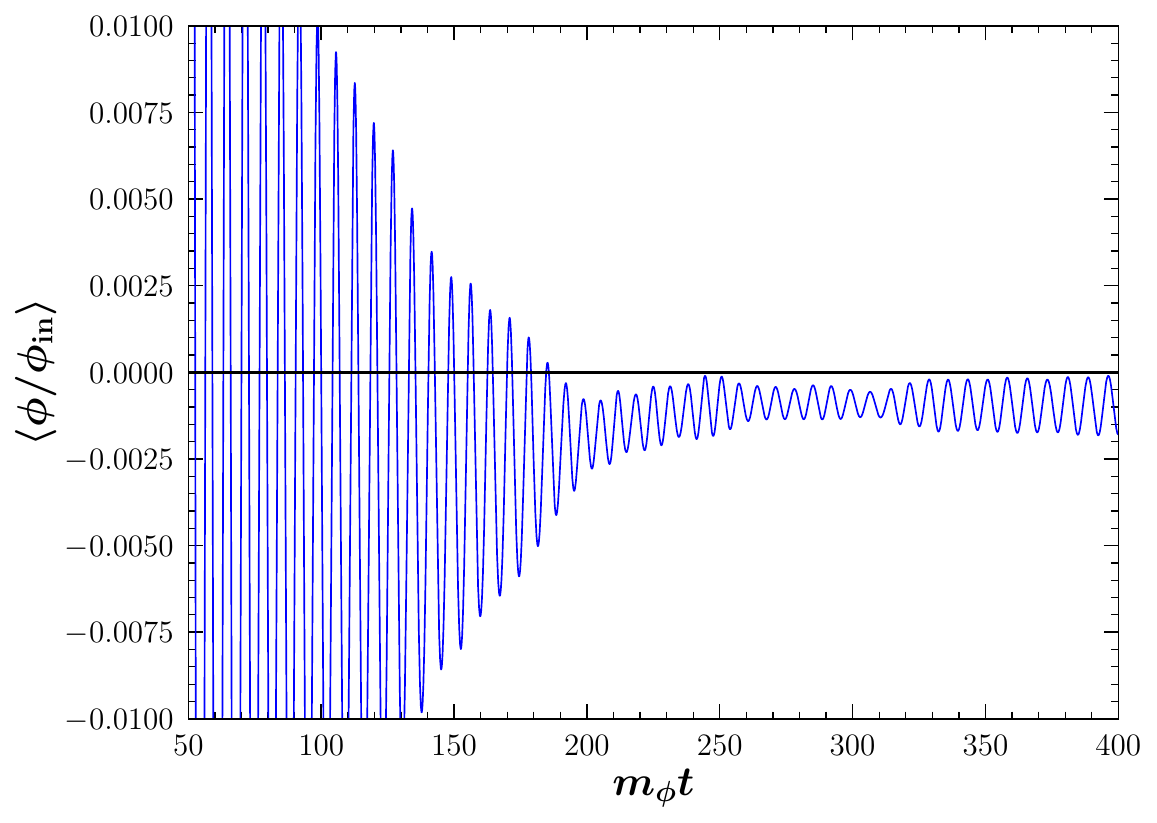}
    \caption{The value of $\phi$ averaged over the simulation box as a function of time for model B. With the introduction of the trilinear term, the back-reaction from the $\chi$ quanta leads to a shift in the global minima of the potential, causing $\phi$ to acquire a non-zero vacuum expectation value (VEV), as evident from the simulation results shown in this figure.}
    \label{fig:enter-label}
\end{figure}

Another interesting feature of model B is the existence of a perturbative decay channel of $\phi$ into pairs of $\chi$ particles with a rate
\beq
\Gamma_{\phi\to 2\chi} = {\sigma^2 \over 8\pi m_\phi} = {q_B^2 m_\phi^3\over 8\pi \phi_{\rm {in}}^2 }\, .
\eeq
where we used the definition of the resonance parameter $q_B\equiv \sigma \phi_{\rm {in}}/m_\phi^2$.
  To achieve efficient parametric resonance, one must choose $q_B>1$. The simulations for the benchmarks presented in the main text were performed using $q_B=100$.
By using $\phi_0\sim  M_{\rm {Pl}}$ and $\sigma \phi_0 / m_\phi^2 =100$, we can eliminate $\sigma$ from the expression of the decay rate, leading to
$\Gamma_{\phi\to 2\chi}\simeq {10^3 m_\phi^3 / M_{\rm {Pl}}^2}$. For $m_\phi\gtrsim10^3$ GeV, $\phi$ decays before BBN, yielding a viable scenario. However, the GW frequency corresponding to such masses would be typically higher than the sensitive range of LIGO/CE/ET. For $m_\phi \lesssim 10^7$ eV, the $\phi$ field has not decayed by today, causing concern of over-closure as for part of the parameter region in model A, which can be alleviated by invoking additional coupling of $\phi$ to sterile neutrinos. For much smaller masses of $m_\phi \lesssim 10^{-13}$ eV, $\phi$ is potentially a DM candidate.


\providecommand{\noopsort}[1]{}\providecommand{\singleletter}[1]{#1}%
\providecommand{\href}[2]{#2}\begingroup\raggedright\endgroup

\end{document}